\documentclass[12pt,preprint]{aastex}

\begin{document}

\def\arcsecpoint{$''\!.$}
\def\arcminpoint{$'\!.$}
\def\deg{$^{\rm o}$}
\def\ltsim{\raisebox{-.5ex}{$\;\stackrel{<}{\sim}\;$}}
\def\gtsim{\raisebox{-.5ex}{$\;\stackrel{>}{\sim}\;$}}
%\journalid{}{}
%\articleid{}{}

%\slugcomment{submitted to {\it Publications of the Astronomical Society of the Pacific }}

\shortauthors{Dunn, et al.}
\shorttitle{Intrinsic Absorption Characterization with $\it FUSE$}

\title{Intrinsic Absorption Properties in Active Galaxies Observed with the Far Ultraviolet Spectroscopic Explorer\altaffilmark{1}}

\author{Jay P. Dunn\altaffilmark{2},
D. Michael Crenshaw\altaffilmark{2},
S. B. Kraemer\altaffilmark{3},
\& M. L. Trippe\altaffilmark{2}}

\altaffiltext{1}{Based on observations made with the NASA-CNES-CSA Far Ultraviolet 
Spectroscopic Explorer. $\it FUSE$ is operated for NASA by the Johns Hopkins University 
under NASA contract NAS5-32985.}

\altaffiltext{2}{Department of Physics and Astronomy, Georgia State University, Atlanta, GA 30303. 
Email: dunn@chara.gsu.edu, crenshaw@chara.gsu.edu, trippe@chara.gsu.edu}

\altaffiltext{3}{Institute for Astrophysics and Computational Sciences, 
Department of Physics, The Catholic University of America, Washington, DC 
20064; kraemer@yancey.gsfc.nasa.gov.}

\begin{abstract}

In a continuing survey of active galactic nuclei observed by the
$\it Far~Ultraviolet~Spectroscopic~Explorer$, we provide a 
deeper analysis of intrinsic absorption features found in 35 objects. 
Our survey is for low-redshift and
moderate-luminosity objects, mostly Seyfert galaxies.
We find a strong correlation between maximum radial velocity and 
luminosity. We also examine the relationships between equivalent 
width (EW), full width at half maximum, velocity and continuum flux. The 
correlation between velocity and luminosity has been explored previously 
by Laor \& Brandt,
but at a significantly higher redshift and heavily weighted by
broad absorption line quasars. We also have 
examined each object with multiple observations for variability in each 
of the aforementioned quantities, and have characterized the variation 
of EW with the continuum flux. In our survey, we find 
that variability of O VI $\lambda$$\lambda$1032, 1038 is less common 
than of the UV doublets of C IV and N V seen at longer wavelengths, 
because the O VI absorption is usually saturated. Lyman $\beta$ 
absorption variability is more frequent. In a target-by-target 
examination we find that broad absorption line and narrow 
absorption line absorbers are related in terms of maximum outflow
velocity and luminosity, and both can be exhibited in similar 
luminosity objects. We also find one object that shows radial velocity 
change, seven objects that show equivalent width variability and two 
objects that show either transverse velocity changes or a change in 
ionization.

\end{abstract}

\keywords{galaxies: Seyfert -- ultraviolet: galaxies}

\section{Introduction}

It is commonly known that intrinsic UV absorption occurs in a large
fraction ($\sim$ 50\%) of Seyfert galaxies. First noticed by Anderson 
\& Kraft (1969), several surveys have been done in recent years 
(Crenshaw et al. 1999; Kriss 2002; and Dunn et al. 2007, hereafter 
Paper I) to
characterize these features. In the first step of our survey we 
examined low-z active galactic nuclei (AGN) (Paper I). Seyfert 
galaxies are moderately luminous, relatively nearby, and mostly spiral 
galaxies, which make up a large portion of our survey, while the 
remainder are low-redshift quasars. We concentrate only on AGN that 
show broad permitted emission lines and narrow emission lines (i.e., 
Seyfert 1's, Kachikian \& Weedman 1974). This is because intrinsic
absorption is only readily detectable if seen against the broad 
emission lines and central continuum source characteristic of 
type 1 objects. Seyfert 2 galaxies have 
only narrow permitted and forbidden lines, and thus intrinsic 
absorption cannot be easily detected. 

Crenshaw et al. (1999) performed a survey of Seyfert galaxies that
had been observed with the Faint Object Spectrograph (FOS) and
investigated the properties of the intrinsic absorption lines.
They found that nearly 60\% of the targets in the survey showed
narrow C IV absorption features (FWHM $<$ 500 km s$^{-1}$) and  
outflow velocities up to 2000 km s$^{-1}$. Also in this survey,
they found a link between warm X-Ray absorbers and UV intrinsic 
absorption, in the sense that Seyfert galaxies that show one also
show the other. 

Broad absorption line (BAL) quasars have been studied even more 
extensively. These objects show lines with much larger velocity 
widths, up to $\sim$ 10,000 km s$^{-1}$ (Weymann et al. 1981, Turnshek 
1988), which appear in $\sim$ 10\% of all quasars (Foltz et al. 
1990). This is a much smaller percentage than the percentage that
exhibit narrow absorption lines only, so the global covering factor is 
an important consideration in determining their nature.

Laor \& Brandt (2002) looked for various trends concerning intrinsic 
absorption in data from the {\it Hubble Space Telescope (HST)} and the 
{\it International Ultraviolet Explorer (IUE)}. Their survey was 
derived from the PG quasar sample (Boroson \& Green 1992). One of the 
more notable trends they found was a dependence of maximum outflow 
velocity on luminosity (v$_{max}$  $\propto$ L$^{0.62 \pm 0.08}$). Their 
plot contains both BAL and narrow absorption lines (NALs). They found 
that the equivalent 
width (EW) also followed an increase with increasing luminosity. This
trend is based on Soft X-Ray Weak Quasars (SXWQs), which show larger
values of EW and v$_{max}$ than any other quasar across the luminosity
span. They furthermore showed that $[$O III$]$ EW and v$_{max}$ are 
correlated. SXWQs showed a significant correlation of 
EW versus v$_{max}$. Of the 56 objects they investigated, 28 showed 
intrinsic UV absorption. This agrees well with the $\sim$60\% frequency 
Crenshaw et al. (1999) found in their lower-luminosity sample and the 
$\sim$50\% frequency we found in Paper I.

For a purely BAL sample, Ganguly et al. (2007) recently used
data from the Sloan Digital Sky Survey (SDSS) to investigate their 
properties.
They conclude that the overall fit between v$_{max}$ and luminosity 
found by Laor \& Brandt applies to their much larger sample of over
5000 objects. In Ganguly \& Brotherton (2008), they compile all intrinsic
absorption features for all known surveys and show that the fit still
holds.

Our knowledge of the variability for intrinsic absorption is limited. 
Variability is important because it is our only link to several key 
physical parameters concerning the AGN. Cromwell \& Weymann (1970) were the 
first to detect variability, in the Seyfert 1 NGC 4151. Later, in 
data from the {\it IUE}, more absorption variability was seen (Crenshaw et 
al. 1999). Nearly all of the variability has been due to changes in 
the ionic column densities, as a result of changes in the ionizing 
flux from the continuum source. However, in four Seyfert galaxies (NGC 
3516, NGC 3783, NGC 4151, and NGC 5548), bulk transverse motion of gas 
across the line of sight was identified as a likely culprit 
(Crenshaw et al. 2003). In the bright Seyfert galaxy NGC 4151, 
both ionization change and bulk transverse motion were detected. Kraemer 
et al. (2006) performed a detailed analysis of the absorption from the
Space Telescope Imaging Spectrograph (STIS) data. 
For the component they dubbed D+Ed, they found a transverse velocity lower 
limit of v$_t$ $\ge$ 2100 km s$^{-1}$.  

NGC 3783 has shown previously to be variable (Gabel et al. 2005, Maran 
et al. 1996). In
1993 February data from STIS,
NGC 3783 showed a spectrum with no narrow intrinsic absorption. By 
1994 January, two obvious absorption features arose for each member of 
the C IV doublet. By 2000 February another pair of absorption lines 
had emerged. Gabel et al. (2003) showed that the radial velocity 
changed as well. In a series of several spectra the radial velocity of 
the line changed in two time spans. They first found a change of 
$\sim$35~km~ s$^{-1}$ and in the second span they saw a change of $\sim$ 55 km 
s$^{-1}$. This is the first well-documented case of a change in radial 
velocity for an intrinsic absorber in a Seyfert galaxy. Transverse 
velocity was also measurable, where they found a lower limit of v$_t$ 
$\ge$ 540 km s$^{-1}$, which is comparable to the radial velocity 
($\sim$$-$450 km s$^{-1}$). Motivated by their findings, we
undertook a survey of O VI and Ly$\beta$ absorption in $\it Far 
Ultraviolet Sepctroscopic Explorer (FUSE)$ 
spectra of low-redshift AGN.

The penultimate goal of our research is to understand the general 
dynamics of mass outflow and to ultimately help complete the picture 
of the 
AGN central engine. Several dynamical scenarios have been offered as 
explanations for the outflow (see Crenshaw et al. 2003 for a review of
the models). One method, Compton heated winds assumes that EUV 
and X-ray photons from the accretion disk are irradiating the gas 
at larger distances and creating a thermal wind (Begelman et al. 
1983). Radiative driving is also a promising method to explain 
fast-moving outflows such as BALs (deKool \& Begelman 1995, Arav et al. 
1994). Another possibility is that magnetic 
fields from the accretion disk are lifting plasma from the disk in 
magnetohydromagnetic (MHD) flows, where the plasma follows a helical 
magnetic field line (Blandford \& Payne 1982). Depending on how far from 
the central core these clouds can follow the field lines, this could also 
be an explanation for the high transverse velocities.

In Paper I we cataloged 35 AGN with intrinsic absorption and 
presented approximate velocities. We found 11 new UV intrinsic 
absorption objects in a sample of 90 nearby AGN observed by $\it 
FUSE$. We calculated the frequency of O VI absorption to be on the order of 
50\%, slightly lower than the value Crenshaw et al. (1999) found for C IV. 
The global covering factor, based on an estimate of the average covering 
factor in the line of sight ($<$C$_f$$>$) was similar to previous 
results of $\sim$0.4. We continue our study of that absorption 
measurements of the absorption lines we found in the 35 objects from Paper I.

\section{Optical Observations and Measurements}

We obtained ground-based optical spectra to obtain redshifts and/or 
estimate black-hole masses and to supplement the {\it FUSE} data for 
three objects: IRAS F22456-5125, MR 2251-178 and WPVS 007. These data are 
readily available in the literature for most of the other objects. Each 
galaxy was observed in both the blue and red optical regions with the 
R-C spectrograph on the Cerro Tololo Inter-American Observatory (CTIO)
1.5-m telescope in Chile. Table 1 chronicles the dates, integration times 
and wavelength coverage for these data. The blue spectra (spanning from 
approximately 3360 to 5440 \AA) were taken using a grating with a 
dispersion of $\sim$1.47 \AA pixel$^{-1}$, providing a resolution of 4.3 
\AA. The red spectra (5652 to 6972 \AA) were taken using a Schott GG495 
filter and a grating with a 1.10 \AA pixel$^{-1}$ dispersion grating giving 
us a resolution of 3.1 \AA. All objects were observed through a long slit 
with a width of 4$^{\prime\prime}$. The stars LTT 4364 and 
Feige 110 were observed with the same settings for the purpose of flux 
calibration. The spectra were reduced and calibrated using standard
IRAF reduction packages for long-slit spectroscopy. We show the 
reduced spectra in Figure 1.

For the source IRAS F22456-5125, which had a redshift estimate listed
on NASA/IPAC Extragalactic Database (NED) of 0.1 (Mason et al. 1995), we 
used our spectra to estimate the redshift based on the positions of lines 
in the blue spectrum. 
The centroids of the emission lines $[$O II$]$ $\lambda$3727, $[$Ne III$]$ 
$\lambda$3869, $[$Ne V$]$ $\lambda$3424 and H$\beta$ were available
in the wavelength coverage of the blue setting and we found an 
average redshift of z=0.1016$\pm$0.0001. The redshifts from 
NED for WPVS 007 and MR-2251-178 are consistent with our measurements.

The optical spectra allowed us to determine mass estimates for the central 
supermassive black hole (SMBH) of these objects via the empirical 
relationships between SMBH mass and the
H$\beta$ FWHM and continuum luminosity at $\lambda$=5100\AA\ 
(or, equivalently the H$\beta$ line luminosity) 
calibrated by Vestergaard and Peterson (2006) using emission-line 
reverberation mapping. We present these masses along with the masses of 
the rest of the sample later in this paper.

\section{UV Observations and Measurement}

We have 105 observations from $\it FUSE$ of the 35 targets 
found with intrinsic absorption from Paper I (see Table 1 from Paper I). 
$\it FUSE$ consists of four telescopes each with four gratings and two detectors 
with a spectral resolution of $\sim$15 km s$^{-1}$, which provides 8 spectra 
per observation covering the wavelength range of 905 \AA\ to 1187 \AA. 
The detectors have two different coatings, LiF and SiC. The detector
with the LiF coating provides a (nearly twice) higher reflectivity than the 
SiC coating (Sahnow 2002). The SiC and Lif detectors do have 
overlapping regions that allow for the spectra to be coadded, weighted by 
exposure time, and scaled to the LiF 1a spectrum. We downloaded the spectra
from MAST and used CalFUSE 3.1 to process the raw data in time-tag mode 
(Dixon et al. 2002). Further details of the reduction process can be 
found in Paper I.

\subsection{Light Curves}

Generating a continuum light curve for $\it FUSE$ data proved to be a 
difficult task due to the nature of the two detectors and redshifted 
broad emission lines. Using the same method as described by Dunn et al. 
(2006), we 
chose 1110 \AA\ in the observed frame as a good position to measure 
the continuum flux. In a 10 \AA\ bin, both detectors contribute to the 
measured flux at that wavelength. In some spectra from either the LiF 
detector or the SiC detector, the target was sometimes out of the 
aperture.  In these cases we did not include the segments. Also, due 
to redshift considerations, we were forced to move the bin when the 
redshift was $\ge$ 0.6. At those particular redshifts the broad 
line would provide a measured flux value too high for the continuum by 
up to 20-30\%. For the majority of the observations, the flux we 
provide is an average flux within the 10 \AA\ bin at 1110 \AA; this 
avoids interstellar medium (ISM) lines (Morton 1991) and geocoronal 
dayglow lines (Feldman 
et al. 2001). The rest of the observations were measured in a 20 \AA\ 
bin taken at 1160 or 1020 \AA, depending on which was closest to and 
unobstructed by the O VI broad line. We converted our fluxes to log 
specific luminosity, and provide flux, log luminosity, bin position, 
and errors in Table 2.

\subsection{Measured Absorption Quantities}

We measured the EW, full width at half-maximum (FWHM) and 
radial velocity centroid with respect to systemic redshift of each 
intrinsic absorption feature identified in Paper I. These 
measurements are listed in Table 3 for the Ly$\beta$, O VI $\lambda$ 
1032 (O VIb) and O VI $\lambda$ 1038 (O VIr) lines. Many features 
were blended or too contaminated by galactic absorption features to 
be measured accurately, thus they have been excluded. Because $\it FUSE$ 
has no onboard 
calibration lamp, CalFUSE is designed to apply a fit to the spectrum 
based on a predetermined wavelength solution (Sahnow 2000). This can 
lead to an inaccuracy of up to 20 km s$^-$$^1$ (Gillmon et al. 2006). 
We measured available ISM lines near our intrinsic absorption features, 
in order to avoid nonlinear calibration effects, to correct the 
velocities. These corrections have already been accounted for in our 
velocities in Table 3.

Errors for EWs were based upon the signal-to-noise (S/N) ratio
of the data along with the standard deviation of three measurements of 
each line with three different continuum estimates. The quoted error
is a sum of these two values. Because velocity centroids, which are
depth weighted, have small standard deviation errors for one line, we 
used the mean standard deviation for the three lines (Ly$\beta$, O VIb, 
O VIr) in each observation. For FWHM we estimated the error as the 
standard deviation of the three measurements per line, similar to EW.

We estimated flux error from the standard deviation within the 10 or 
20 \AA\ bin. This is typically an overestimate of the error due to
oversampling, similar to the case for IUE data (Dunn et al. 2006).
We find that we overestimate by nearly a factor of 2.5
by examining spectra close in time ($\lesssim$ 1 day) for like 
objects. Because most of these objects are Seyfert galaxies, the flux 
level is not likely to change significantly in the span of $\lesssim$ 
1 day. We therefore scaled our flux errors by this amount.

Our listed redshifts for the objects are provided by NED. We chose
them based on the recommended most accurate value, preferably from 
HI-21 cm measurements. However, not all objects have very accurate 
values for the redshift, such as IRASF 22456-5125 ($\oint$2). Some 
of the redshifts are from narrow emission lines, which are often 
blue shifted with respect to HI-21 cm values (Crenshaw et al. 1999). 
This may explain the slightly positive radial velocities found for 
some components in a few objects.

In several cases we found that of the three lines we examined, only 
one or two of the lines were measurable. This is due to either poor 
signal-to-noise, Ly$\beta$ being too weak, or heavy blending with
ISM features.

\section{Correlations}

All of our correlations involving EW, FWHM and/or luminosity show 
measurements from each absorption component in each observation, 
because many of these quantities can be variable over time. In order
to inspect our plots for correlations we simply examine by eye the
data for either a direct trend or for an envelope fit to the data, as
in Laor \& Brandt (2002).

The plots containing velocity measurements are averaged over time due 
to the lack of significant variability in velocity space. Also, error
bars for many of the points are contained within the plot symbol. The 
most tantalizing correlation we have found 
is that of velocity with luminosity. While this was seen in C IV by 
Laor \& Brandt (2002), their objects were mostly quasars with high 
luminosities and redshifts beyond 1.5, which was our redshift cutoff 
on the high end. We share four objects in common. We plot in Figure 2 
all of our velocities measured for each object and each component. In 
general as the luminosity increases, it appears that the maximum 
velocity at each luminosity does as well. This confirms the trend 
that Laor \& Brandt found for higher-luminosity objects and shows
that the trend also holds for lower-luminosity objects such as Seyfert
galaxies. To confirm this further, we converted the magnitudes
provided by Laor \& Brandt to log L$_{\lambda}$ by using the four
targets we share and found a scale factor of 1.7. We plot the
Laor \& Brandt observations along with our own measurements in Figure
3 and see that the two data sets follow remarkably similar trends. The 
relation given by Laor \& Brandt is v$_{max}$  $\propto$L$^{0.62 \pm 0.08}$.
While we do not calculate this, we can see by Figure 3 that all of 
our points fall beneath this maximum velocity trend. 

Ganguly \& Brotherton (2008) show a similar plot. They have plotted 
the Laor \& Brandt (2002) data along with their SDSS data from Ganguly 
et al. (2007), and data from several other sources (references within). 
The plot shows that for all available data the trend from Laor \& 
Brandt holds for all luminosities and velocities.

Along with the measured black hole-masses from the optical data,
we compiled all of the known masses for our survey objects in
Table 4 (we exclude those few without the necessary measured
quantities) and calculated the Eddington ratio. For all objects
referencing Peterson et al. (2004), the masses are from reverberation
mapping; all others were calculated using the relationship found by
Kaspi et al. 2000 and calibrated by Vestergaard \& Peterson (2006):
\begin{equation}
M_{BH} = log\left[\left(\frac{FWHM(H \beta)}{1000\;km\;s^{-1}}\right)^2\left(\frac{\lambda\;L_{5100}}{10^{43}\;ergs\;s^{-1}}\right)^{0.5}\right]
\end{equation}
using the FWHM of the broad H$\beta$ emission line and the
luminosity at 5100 \AA. We find our value for
L$_{Bo}l$=9.8$\lambda$L$_{5100}$ from McLure \& Dunlop (2004).

We plot the maximum velocity, as before with luminosity, against the
Eddington ratio (L/L$_{Edd}$)in Figure 4. There appears to be a weak trend of
maximum velocity the with Eddington ratio. It should be noted that 
the two objects with the highest velocities have low ratios, 
and the two objects with a high ratio have low velocities. In order to 
complete the picture we would require more data and a wider range 
of objects spanning the range of ratios as most of our targets are 
moderate Eddington accreting objects. 

We examined how the EW relates to the specific luminosity 
in our sample, which we show in Figure 5. Overall there appears to be 
no trend. This contradicts the findings of Laor \& Brandt (2002) for
C IV lines, but this could be a simple case of saturation affecting
our EW measurements. The only notable feature in the plot is the lack 
of points for low luminosity and 
low EW. The paucity of points is most likely a selection effect 
of low signal-to-noise for weak lines in objects with low-continuum 
fluxes. It should be noted that, some of our measured EWs are 
measured over what have been seen, in less saturated UV absorption 
lines, to be multiple components (e.g., NGC 4151 in Kraemer et al. 
2005). Due to the heavy blending and saturation in O VI, we simply
measure across blends as one component. We have removed any 
measurements of lines in spectra with a luminosity signal-to-noise 
of $<$ 30 and measurements from spectra that are saturated with 
multiple components, and still find that there is no trend. 

We additionally investigated the relation between EW and maximum 
velocity (not shown). Again, there appears to be very little trend. 
We also examined the relation between FWHM and 
the specific luminosity, as shown in Figure 6. There appears to be no 
relation between FWHM and luminosity. However, as seen with EW, there 
does appear to be a lack of absorption for O VI at low luminosities
and small FWHMs.

\section{Variability}

Our survey is the first to date with a significant number of objects 
with multiple spectra to provide enough data for a time series 
analysis of intrinsic absorption variability in the far ultraviolet. 
Of the 35 targets found in the Paper I survey, 22 of the
objects have multiple observations. Using the EWs,
velocities and FWHMs we provided in Table 3, we examined a time
series of each along with the corresponding fluxes for
each observation from Table 2. Along with the times series data,
we plotted the spectra for each object in chronological
order to see visually how the absorption features changed.
While most objects showed no real trends between EW, FWHM, or
velocity vs. continuum flux, we found a handful of cases that do.

Mrk 79 is one case of a likely variation. Ly$\beta$ is clearly
visible in the observations in Figure 7. Component 1 of O VIb is a 
broad feature due to blending with ISM lines. O VIr is a narrow, most 
likely saturated line. The variability for Mrk 79 is seen in the O VIr 
line. O VIr changes in EW for component 1, but does not appear to be 
correlated with continuum luminosity in Figure 8.

NGC 4151 has broad, saturated absorption troughs.
These troughs are comprised of several blended components seen
as individual features in STIS data (Kraemer et al. 2001). Because
they are blended, it makes measuring individual EWs, FWHMs and velocities 
difficult. 
One interesting point is that in the first two observations, seen in 
Figure 9, the broad trough shows some structure, but in the subsequent 
observations, they blend too much to discern any one particular 
absorber. As shown in Kraemer et al. (2001), one particular component, 
labeled D$^{\prime}$, only appears in weak flux states. In 
Table 2 we see that the earlier observations are weaker in flux, 
and that the D$^{\prime}$ component flattens the region between 
$\sim$1029 and 1032 \AA, which is consistent with the interpretation of 
Kraemer et al.

Mrk 817 is another object that shows evidence of variability. One of 
the labels in Dunn et al. (2007) is incorrect; the Lyman $\beta$ 
line is weak, and the line marked component 1 at 1064 \AA\ is H$_2$. 
In Figure 10, we see that Mrk 817 shows in O VIb that component 
1 evolves from two subcomponents to a smoother single component. It 
is possible that these subcomponents are hidden within the 
artificially broadened O VIr line due to galactic Fe II $\lambda$ 
1055. Lyman $\beta$ is not visible most likely due to high 
ionization. This is confirmed by STIS spectra that show weak 
absorption in Lyman $\alpha$ (Penton, Stocke \& Shull 2002). The EW
for the original lines shows a correlation with the decreasing 
continuum flux (Figure 11), and FWHM shows the same. While this is
a 2$\sigma$ detection in O VIb, the similar behavior of all three lines 
leads us to believe that the change is real. The velocity 
seems to drop for Mrk 817 showing a significant decrease between the 
first observation and the last. All of this suggests two blended 
components in the original observation, and that the higher-velocity 
component decreased with time.

Mrk 817 also demonstrates the appearance of a previously unseen
line. At $\sim$ 1052 \AA\ and $\sim$ 1058 \AA\ we see that in the 
first spectrum there is no significant sign of an absorption feature. 
However in the third spectrum there appears to be a slightly depressed 
broad feature, and in the final spectrum the broad line is clear with 
two O~VI features on either side at the appropriate locations in velocity
space v$_r$=3700 \& 3430 km s$^{-1}$. However, we cannot rule out the 
possibility that the new component is present in the previous 
observations but not detectable due to the lower S/N. 

In Figure 12, we present the available spectra for Mrk 279. Figure 
13 shows the relation of EW and continuum flux for Mrk 279, which 
shows a weak trend. In both O VI lines the EW increases with 
decreasing flux. For Ly$\beta$ the relationship with flux is much 
flatter. During the last three observations we see the flux drop 
by $\sim$60\% and then increase by $\sim$50\%, while the EW for all three
lines shows a chronological decrease. We see that O VIb varies more than
O VIr in Mrk 279. This is most likely due to contamination in both of the
O VI lines from ISM and H$_2$. While these lines are heavily contaminated
(see Paper I), we can still measure variation as the ISM 
lines should experience no change.

Mrk 290 (Figure 14) provides a case for a weak anti-correlation between 
EW and flux. 
In this object the EWs for both members of the O VI doublet 
$\lambda$$\lambda$ 1032 and 1038 
lines show a slight decrease with a significant increase in continuum 
flux over a 1200 day interval, and a moderate decrease in the Ly $\beta$
width, as shown in Figure 15. 

The variability in Mrk 509 was complex, as shown in Figure 16. What we 
measured for component 1, can be seen in Lyman $\beta$ to be at least two
subcomponents in two of our three spectra. What we call component 
2 can be seen in Lyman $\beta$ to be comprised of at least two or three 
subcomponents. Kraemer et al. (2003) confirm this by identifying 8 
components in spectra from FOS and STIS, which are blended together 
with the two main components we see in the $\it FUSE$ data. The 
components labeled 1-3 by Kraemer et al. are subcomponents that are 
contained in our measured component 1, while those labeled 4-8 by 
Kraemer et al. make up our measured component 2. Subcomponents 3 and 6 
in Lyman $\beta$, labeled by Kraemer et al. appear to vary. 
Subcomponent 3 shows an decrease in depth while subcomponent 6 
increases in depth over time. We discuss the depth further in the next 
section. Our measured component 1 shows no change in EW or velocity 
over time (Figure 17), but our measured component 2 shows a slight 
increase in Ly$\beta$ (Figure 18).

Figure 19 shows the spectra of PG 0804+761, which exhibits evidence of 
radial velocity change. This change is seen in all three lines as a 
centroid shift to more negative velocities, while the ISM lines 
remain stationary. We plot these centroid velocities vs. time (Figure 
20). PG 0804+761 has a poor estimate of redshift, similar to 
IRAS~F22456-5125. Thus the absolute values of velocity reflect 
redshifted lines, but this might change with a more accurate redshift.

In the object NGC 3783, we find variability of EW in the Ly $\beta$ line.
Figure 21 shows the time evolution of the spectrum, and Figure 22 shows 
the relation between flux and EW and velocity over time. The EWs for 
O~VI $\lambda$1038 and O~VI $\lambda$1032 show no correlation,
across the dashed average EW line. Ly$\beta$ decreases from low flux 
to high flux. However, this is evidently a blended combination of two 
lines labeled 2 and 3 in a paper by Gabel et al. (2003). Thus it is
difficult to determine if one or both lines are responding to continuum
change. In Figure 21, we see that components 1 and 4 from Gabel et al.
(2003) are visible in Ly $\beta$, and are hard to discern in O VI $\lambda$
1032 due to heavy contamination from C II and H$_2$. For O VI $\lambda$ 1038
there are two H$_2$ lines nearby, but the weak component 4 is not visible.

\section{Discussion}

In the Kraemer et al. (2005, 2006) study of the variability in NGC~4151, 
they found some cases where the EWs of lines were anti-correlated with 
the continuum flux level, while others showed no relation to the flux.
We see similar events in only a few objects such as Mrk 817, where the 
O~VI lines seem to follow the continuum, and Mrk 79, where the EWs 
vary but the continuum stays relatively constant. The paucity of O~VI
variability suggests that O~VI, which is usually predicted to be very 
strong in absorber models, is saturated in most cases. It should also
be noted that in many cases there is a lack of change in Ly$\beta$, 
this is presumably also due to saturation. The changes we do
see in EW must be due to one of the three events: a covering factor change
in the line of sight, a change in ionization, or a total column change
(characterized by N$_H$) due for example, to bulk motion of gas into 
our line of sight.

\subsection{Potential Cases of Bulk Motion}

There are three targets in our survey that show changes in the 
absorption are the best candidates for bulk motion of gas across the
line of sight with a transverse velocity (v$_t$), as described below. 
These three objects are Mrk 79, Mrk 817 and Mrk 509.

Mrk 79 is an interesting target because while the EW of component 1 
is variable, the continuum does not change. This lends itself to a
case of changing the total column, most likely due to bulk motion. We can 
calculate the transverse velocity by using the measured broad line 
diameter from Peterson et al. (1998) of 36 light days and a time
interval between observations ($\Delta$t) of 83.7 days the absorber 
would be moving at a transverse velocity $\ge$126,000 km s$^{-1}$. 
Given the average radial velocity of the absorber v$_r$$\approx$300 
km s$^{-1}$, the estimate for the transverse velocity is inordinately 
high. We will tackle further explanations in the next section.

In the case of Mrk 817, if we assume that the appearance of the new 
absorption components (2 and 3) are due to transverse motion of 
material into the line of sight, we can follow the methods used by 
Gabel et al. (2005) for NGC 3783. Assuming the O~VI doublet is 
not saturated we can use the covering factor in the line of sight 
for O VI at the conclusion of the variability event with the equation:
\begin{equation}
C_f = \frac{I_1^2 - 2I_1 + 1}{I_2 - 2I_1 + 1}
\end{equation}
(Crenshaw et al. 2003), where I$_1$ is the intensity at the core of the 
stronger oscillator strength line and I$_2$ is for the weaker line. This 
number can be skewed an by additional flux from the Narrow Emission Line 
Region (NELR) (Kraemer et al. 2002, Arav et al. 2002, Crenshaw et al. 
2003), but for Mrk 817 the narrow emission lines are not prominent, and 
we find C$_f$ = 0.70. If we compare this to the line-of-sight covering 
factor of 0.51 and 0.33 for the blue and red lines respectively we verify
that the lines are not saturated due to the fact that as lines approach 
saturation in a doublet, the line-of-sight covering factor approaches the
doublet covering factor. The transverse size of the absorbing cloud is 
d=d$_{BLR}$$\times$$\sqrt{C_f}$ (Crenshaw et al. 2003). For Mrk 817, the 
BLR is 31 light-days across (Peterson et al. 1998). The absorption 
appeared over $\sim$310 days, which gives a lower limit to the transverse 
velocity of $\sim$25,100 km s$^{-1}$. This is exceptionally large 
compared to the radial velocity of 3700 and 3430 km s$^{-1}$ for the two 
new components. Thus, either the components were buried in the noise
in the earlier observations or we are witnessing their appearance due to 
ionization changes.

Mrk 509 was shown in Peterson et al. (1998) to have a BLR size of 159
light days. If we assume that subcomponents 3 and 6 are saturated in 
Ly$\beta$, then we can calculate the change in covering factor (C$_f$)
from the residual intensities in the cores (C$_f$ = 1-I$_r$). In the 
case of subcomponent 6, we measured the core and found average 
covering factors C$_f$=0.78 and 0.96 for observations 1 and 3 
($\Delta$t=307.3 days). For subcomponent 3, we find covering factors 
C$_f$=0.93 and 0.77 for observations 2 and 3 ($\Delta$t=304.0 days). 
Subcomponent 6 therefore, showed an increase in covering factor by 0.18 
while subcomponent 3 showed a decrease by 0.16. Based on the changes 
in covering factors, from our previous equation we can find lower limits 
for transverse velocity for these of 65,600 km s$^{-1}$ for subcomponent 
6 and 125,200 km s$^{-1}$ for subcomponent 3. This seems extremely high 
for a transverse velocity when compared to the radial velocities of 
$\sim$300 km s$^{-1}$. Therefore, the changes in the depths of these two
subcomponents are almost certainly due to ionic column variation and 
not covering factor changes.

Thus, even in the most favorable cases, we find no evidence for EW changes
due to bulk motion. Because there is no evidence that transverse velocities
of UV absorbers can reach these inordinately high values (Crenshaw et al. 
2003), the only viable alternative is ionic column changes due to changes 
in the ionizing continuum.

\subsection{Ionization Changes}

The three objects we have ruled out bulk motion for and the other six 
objects that showed signs of absorption variability all seem to be the
result of changes in the ionizing flux. Mrk 279 and Mrk 290 showed an 
anticorrelation between ionization parameter (U) and O VI EW, while 
NGC 3783 showed anticorrelation in Ly$\beta$. For Mrk 279 and 
Mrk 290 this means that the ionization state is high, and NGC 3783 is
indeterminate due to no detectable change in O VI. Mrk 817, component 1,
is the only case where the EW for O VI is correlated with the flux, which 
means that U is relatively low. For NGC 4151 we see the D$^{\prime}$ 
component decrease with increasing flux, thus it is another high-U 
component. PG 0804+764 showed correlation between flux and EW for both
the O VIb line and Ly$\beta$. This object stands out because it is the 
only object in the survey that showed an unexplainable change in 
radial velocity.

Mrk 79 showed a change in EW and no change in flux. Because we only have 
three observations, we believe what we are seeing a change in ionization 
with a time delay. Our observations do not accurately sample the 
lightcurve and there must be a delay between the EW response and the 
continuum, due to low densities and/or large distances of the absorbers
from the continuum source (Crenshaw et al. 2003).

The appearance of the absorption components 2 and 3 in Mrk 817 is most 
likely due to a change in ionization. The light curve shown at the 
bottom of Fig. 11, shows that in the last observation where the two 
components are visible, the flux level had decreased significantly. 
The appearance of these lines may therefore be due to a decrease in 
ionizing flux. We can also use the possible disappearing subcomponent 
in component 1 to place limitations on the absorption systems. As the 
continuum flux decreases the line weakens and vanishes/blends with 
the component at $\sim$1049.2 \AA, and could be the explanation of 
the EW variations we noted in $\oint$5. In order to place these 
constraints, we need CLOUDY modeling of the absorbers.

Mrk 279 presents a similar situation. In Figure 13 we have a range
of $\sim$58 days where the continuum flux drops and rises. The EW for
the absorbers respond by decreasing. We could estimate a lower
limit to the distance and density provided a model of the absorber 
and assuming the response is complete after 58 days.
Ideally, we would need a monitoring campaign with several observations
for an event such as this to fully explain the variations.

For the component Kraemer et al. (2002) labeled 3 in Mrk 509, we can 
assume that the ionizing flux increased and  
the column changed. We know from Kraemer et al. (2003) that the ionization
parameter (U) for Mrk 509 in subcomponent 3 is 0.03. 
Given the change in time, $\Delta$t = 2.6$\times$10$^{7}$~s, we can use the
equation as seen in Krolik \& Kriss (1995):
\begin{equation}
t_{ion} = \frac{h \nu}{F_{ion}<\sigma_{ion}>},
\end{equation}
where $\nu$ is the transition frequency, F$_{ion}$ is the ionizing flux
at the cloud front and $<$$\sigma$$_{ion}$$>$ is the ionization cross section
(Osterbrock 1989). Using our known limit on the timescale, and a calculated 
value of L$_{ion}$=2.1$\times$10$^{44}$ ergs s${-1}$, we can get a limit on 
the distance from the source to the cloud from  
F$_{ion}$=L$_{ion}$/4$\pi$d$^2$. That gives us the lower limit to the 
distance of 14 pc.

Finally, we can estimate the density of the absorbing cloud assuming a change in
ionization, from the ionization parameter:
\begin{equation}
U = \frac{\int\frac{L_\nu d\nu}{h\nu}}{4 \pi c r^2 n_e}
\end{equation}
For our subcomponent 3 we find n$_e$$\le$1.38$\times$10$^3$ cm$^{-3}$.

\subsection{NALs and BALs}

From both the Paper I survey and the Laor \& Brandt surveys, one
object, PG 1411+442, appeared in both surveys with BAL and NAL 
signatures. The NALs appear superimposed on the BAL in both 
members of the O VI doublet, as seen in Paper I. This evidence 
tied with the Laor \& Brandt and our relationship from section 
$\oint$4, shows that both BALs and NALs are related and follow the 
same trends and that they both may exist simultaneously in an object.

Also from our previous survey, WPVS 007
was omitted from the search for NALs due to the discovery of a newly
appearing BAL by Leighly et al. (2007, in prep). We have reexamined
this object and found that the NALs seen by Crenshaw et al. (1999)
are still visible in the {\it FUSE} spectrum. The
NALs for O VI and Ly $\beta$ are buried in the BAL and have a velocity of
-399 $\pm$ 10 km s$^{-1}$ seen in Figure 23. This matches the velocity
Crenshaw et al. (1999) found of -390 km s$^{-1}$.

We can evaluate the configuration of the NAL and the appearing BAL in 
WPVS 007. The time difference between the $\it FUSE$ spectrum and the 
last FOS spectrum is 2655 days. If we use the relationship from 
Vestergaard \& Peterson (2006):
\begin{equation}
R \propto L_\nu(\lambda 5100)^{0.50\pm0.06},
\end{equation}
with a BLR size that has been estimated from reverberation mapping, we 
find that the BLR size for WPVS 007 is $\sim$23 light days across. This 
gives a lower limit on the transverse velocity of the BAL of 2565 km 
s$^{-1}$. The NAL however is still apparent in the O VI doublet and Lyman 
$\beta$ lines. This leaves us with two situations. Either the NAL is 
interior to the BAL and is still receiving sufficient ionizing flux to 
maintain the absorption, or the NAL is exterior to the BAL and is no 
longer receiving ionizing flux due to the BAL absorption. If the latter 
is true, then the recombination timescale is larger than the 2655 days
between observations, which could be used to estimate a density and 
distance provided photoionization modeling.

\section{Conclusions}

Our goal was to explore the Dunn et al. (2007) $\it FUSE$ survey in
depth for correlations and variability. We sought to characterize the
intrinsic absorption seen in nearly 50\% of all nearby active galaxies
in order to help characterize the outflow from the nuclear engine.

As shown by Laor \& Brandt (2002), there is a relation between the
maximum outflow velocity and the luminosity. Their survey contained
moderate redshift objects and BAL quasars, while our survey was for 
smaller redshifts and luminosities. However, as shown previously, our 
relation between v$_{max}$ and L is consistent with theirs. In both 
our surveys there is a distinct upper limit on the maximum velocity 
dependent upon the luminosity of the object. This leads us to believe that
both classical narrow intrinsic absorption lines and broad absorption
lines are governed by the same relationship, thus they may be related,
in the sense that they may share a common driving mechanism. Due to 
their simultaneous presence in WPVS 007 and PG 1411+442 however, we 
feel that the nature of their origins may be different.

Laor \& Brandt (2002) also found for higher redshift galaxies that 
EW \& luminosity are correlated for C IV. We examined correlations 
between the observed lines and the AGN properties. We found no link 
between the EW and the luminosity. Because EW
increases with column density of the particular ion in question, 
it would seem that an increase in flux would increase the O VI EW in 
cases where the ionization parameter is low and vise versa for a 
highly ionized cloud, at least until the line reaches heavy 
saturation. We do not see this trend for either O VI or Ly$\beta$. 
This is most likely due to the heavy saturation in O VI and even 
Ly$\beta$ in many cases while C IV and N V do not saturate as easily.

It has been seen in the past that some objects show bulk transverse
motion along with the more easily detectable radial motion (Crenshaw
et al. 2003). Of the 35 objects available we have 22 with more than 
one observation. Two of these objects have only one usable observation, 
leaving 20 total useful data sets. In the survey we find 8 that show 
change. The frequency of variability events seems to be low, in the 
$\it FUSE$ data. With 22 available objects, only 8 show variability. 
This could mean the events are rare, that only weak absorption 
is subject to this change and/or the sample is biased by small number 
statistics, or the O VI saturation makes detecting variability 
difficult.

None of our objects with EW changes can be explained with transverse 
bulk motion. We have 7 that can be explained with ionization changes. 
In the case of Mrk 509 with modeling of the subcomponents, we can find 
a distance and density (n$_H$) of the third subcomponent of 0.3 pc and 
3.8$\times$10$^{8}$ cm$^{-1}$, respectively. Also, we find one object 
that shows evidence for radial velocity changes (PG 0804+761); we 
currently have no explanation for this.

We find in Mrk 817 that the absorption appears in relatively weak lines, 
and furthermore even the changes in the absorption features for Mrk 509 
are in relatively weak lines. This is the case for most transversely 
moving clouds seen in the past. With the low signal-to-noise $\it FUSE$ 
data it may be that these weak lines are swamped by the surrounding 
noise and only strong absorption features are noticeable, or it may be 
that bulk transverse motion is not a common event. Given that these 
changes are in weak absorbers, it suggests that strong absorbers appear 
more static in the far UV than weak absorbers due to saturation and 
blending in many of the lines.

\acknowledgments

This research has made use of the NASA/IPAC Extragalactic Database (NED)
which is operated by the Jet Propulsion Laboratory, California Institute of
Technology, under contract with the National Aeronautics and Space
Administration. This research has also made use of NASA's Astrophysics Data
System Abstract Service. We acknowledge support of this research under NASA
grants NNG05GC55G, NNG06G185G and NAG5-13109.

\clearpage

\figcaption[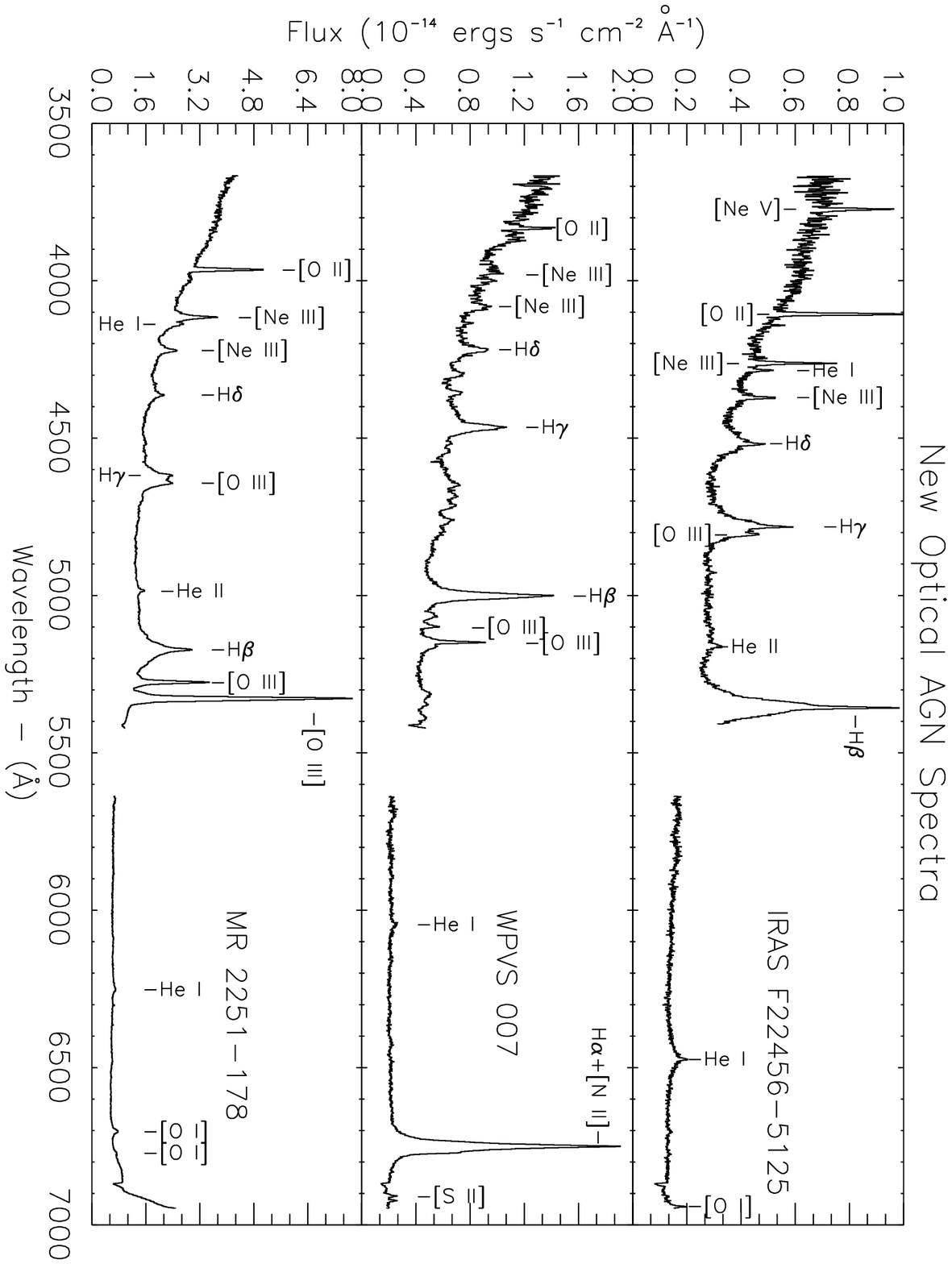]{New spectra taken at CTIO, plotted in the observed frame.}

\figcaption[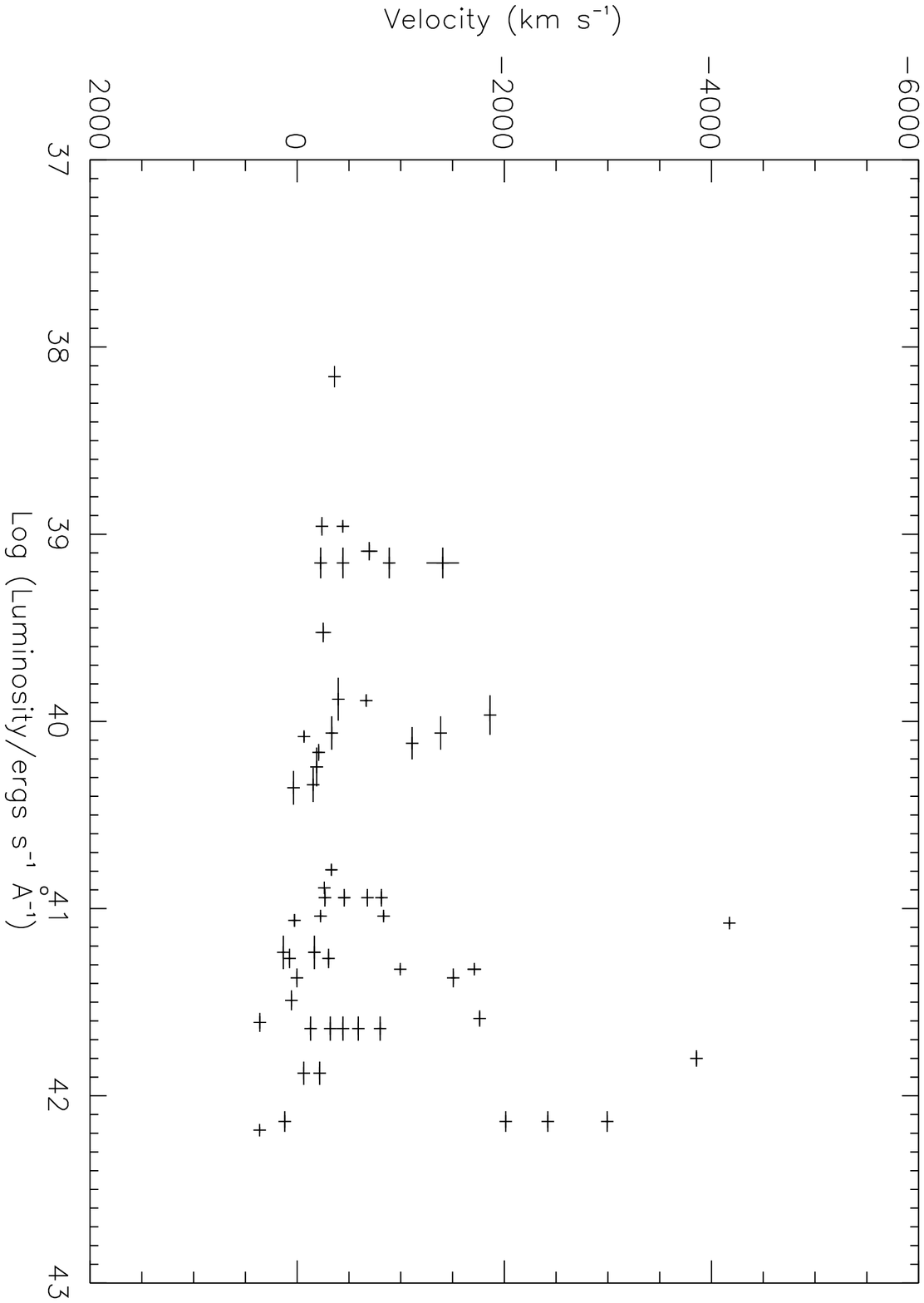]{The relation between luminosity and velocity using only data from $\it FUSE$.}

\figcaption[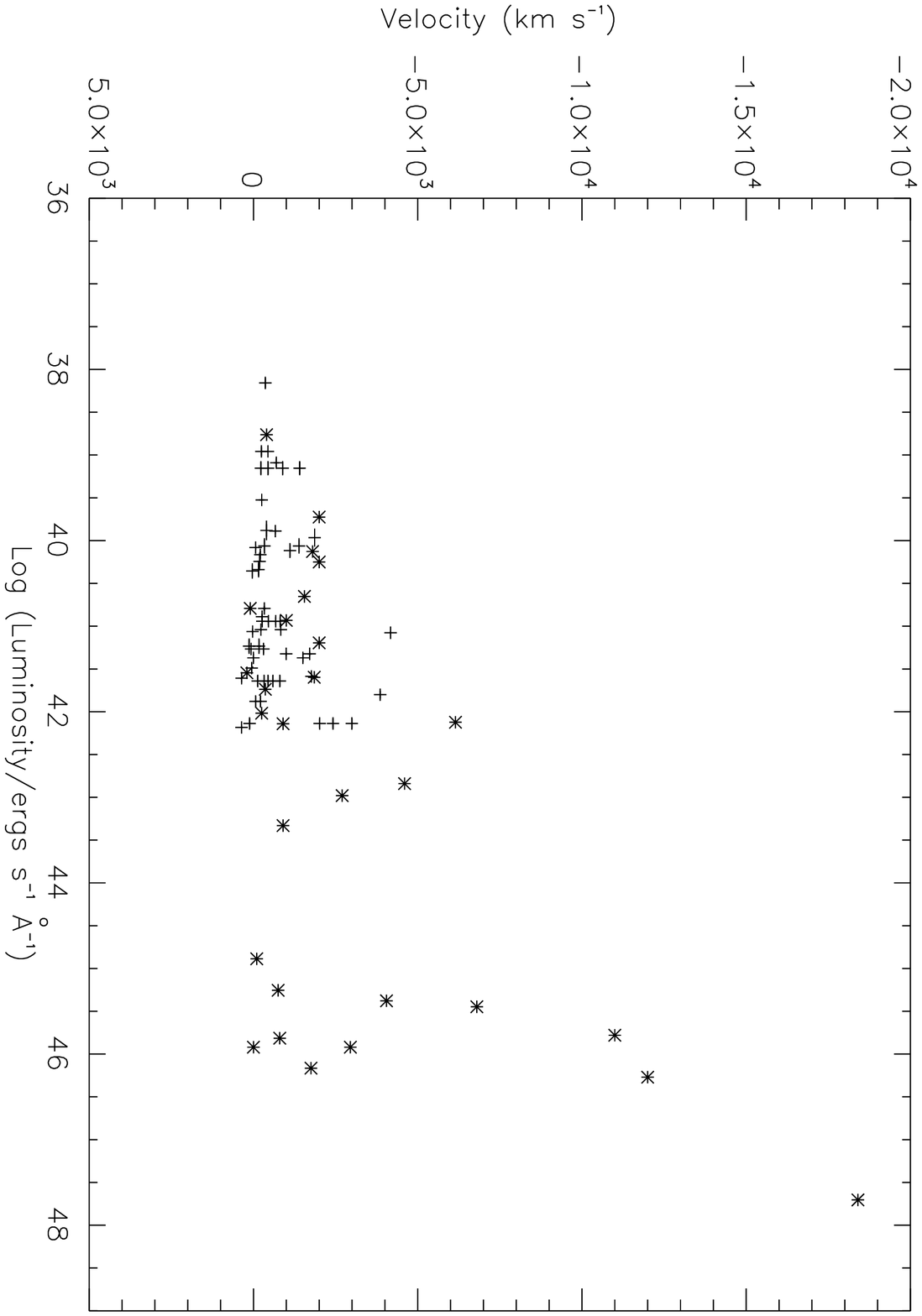]{Extended plot of luminosity and velocity using data measured by Laor \& Brandt (2002). 
Crosses are data from our $\it FUSE$ observations and asterisks are from the Laor \& Brandt (2002) survey.}

\figcaption[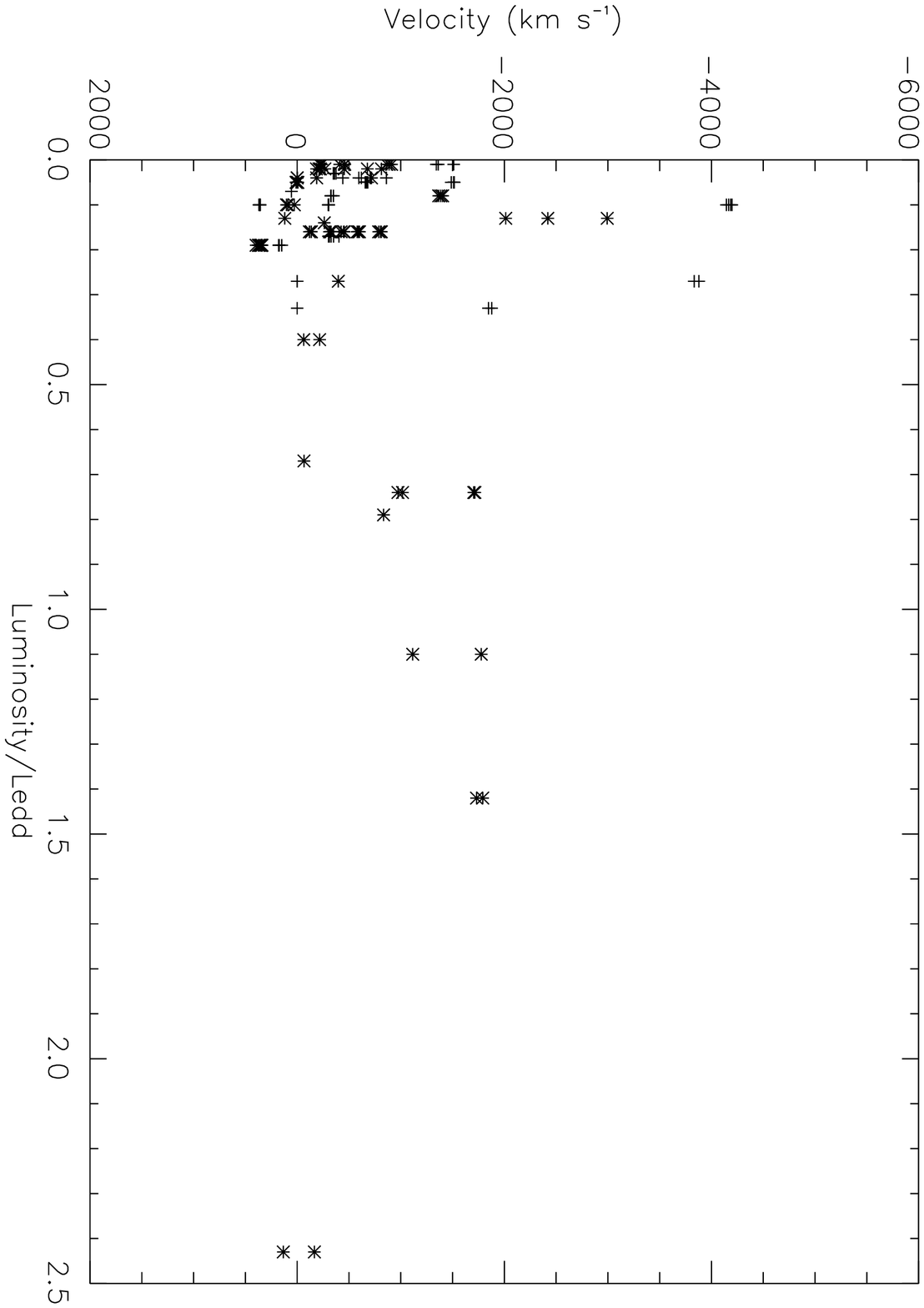]{Relationship between Eddington Ratio and maximum outflow velocity. Crosses are
objects with reverberation mapping.}

\figcaption[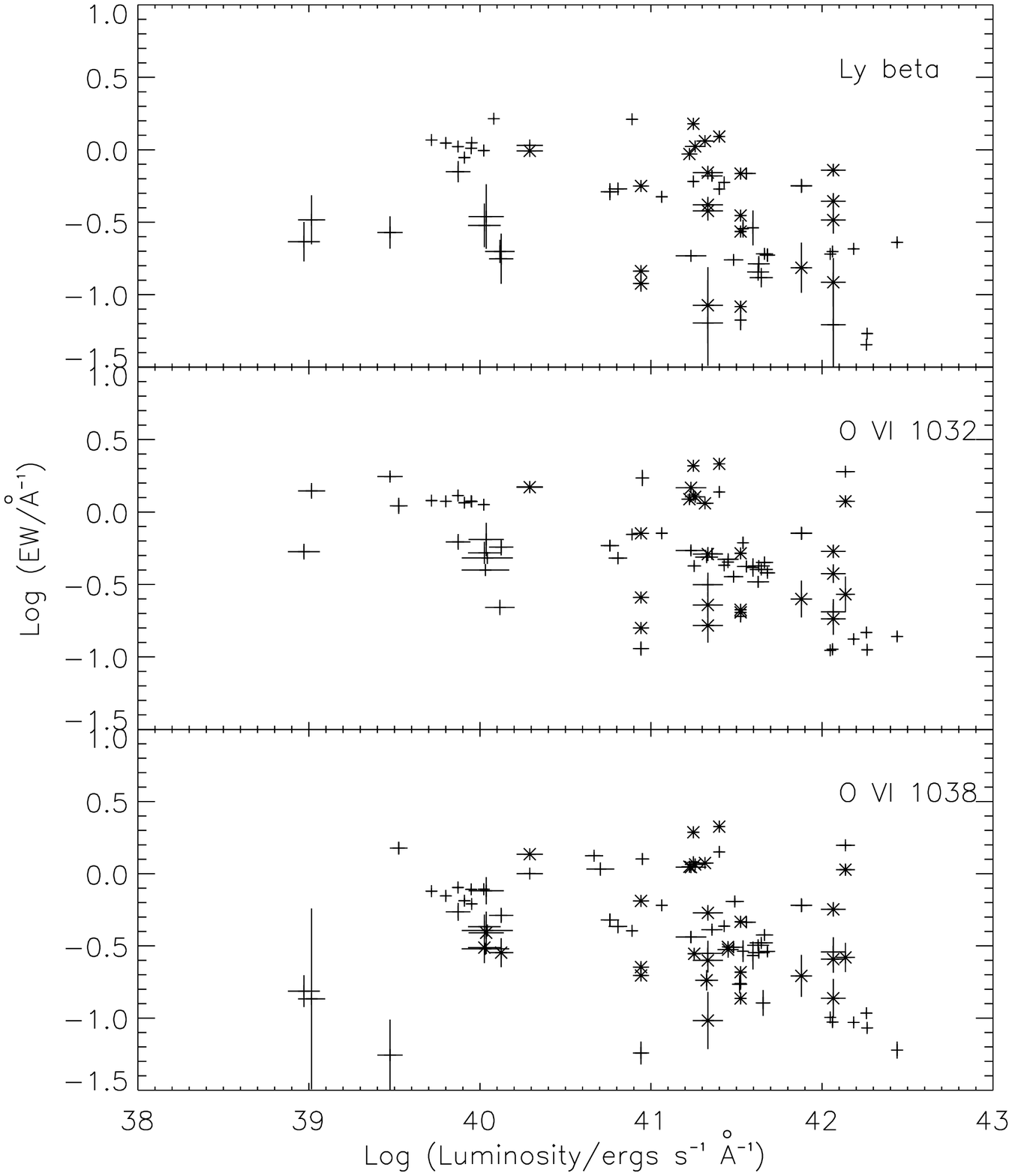]{EW vs. specific luminosity for our sample. The top plot is for O VI $\lambda$1032, 
the middle is for O VI $\lambda$1037 and the bottom plot is for Ly $\beta$. Plus symbols are components 
labeled 1 from Paper I and asterisks any component other than the first.}

\figcaption[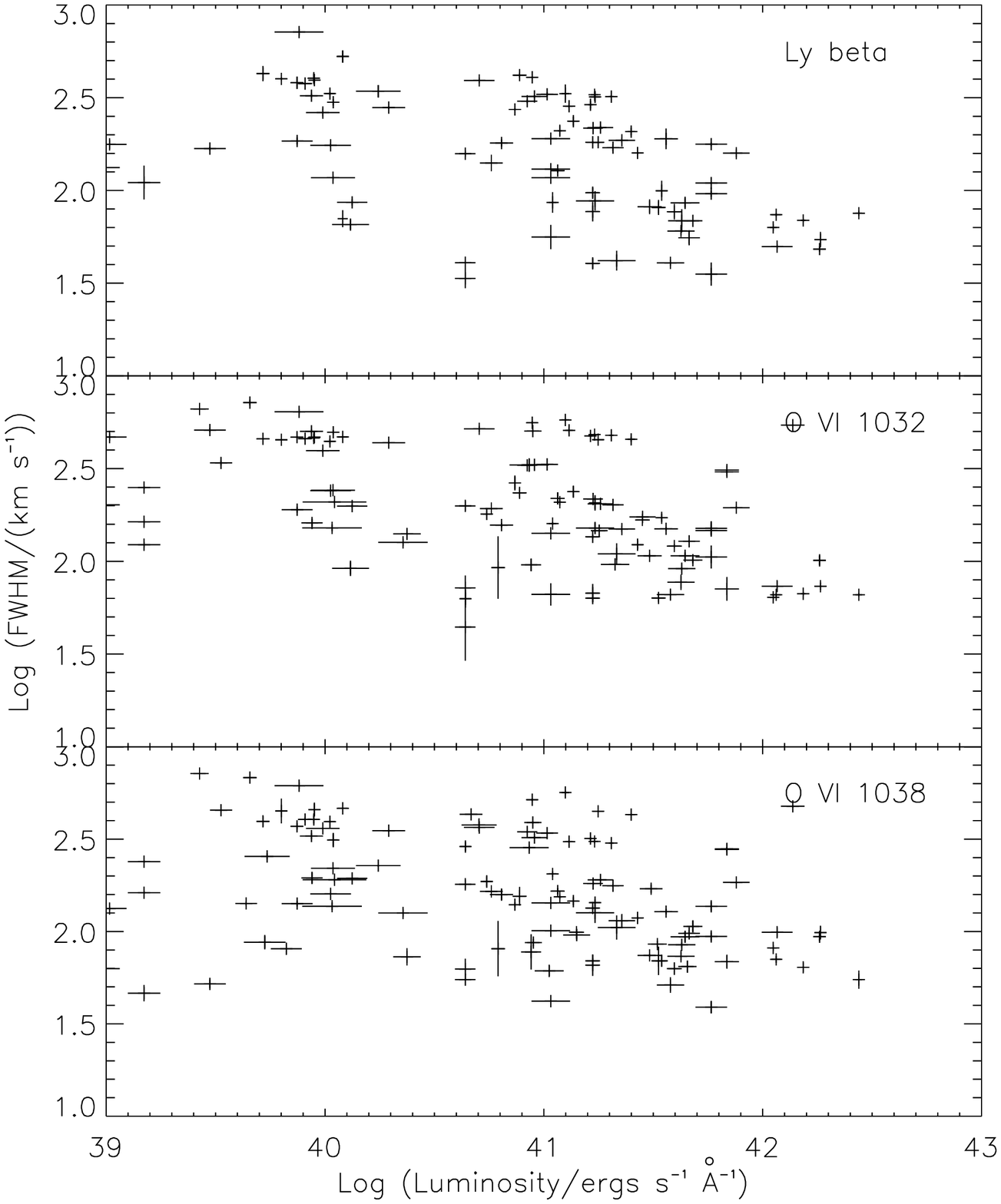]{FWHM vs.specific luminosity. The top plot is for O VI $\lambda$1032, the middle 
is for O VI $\lambda$1037 and the bottom plot is for Ly $\beta$.}

\figcaption[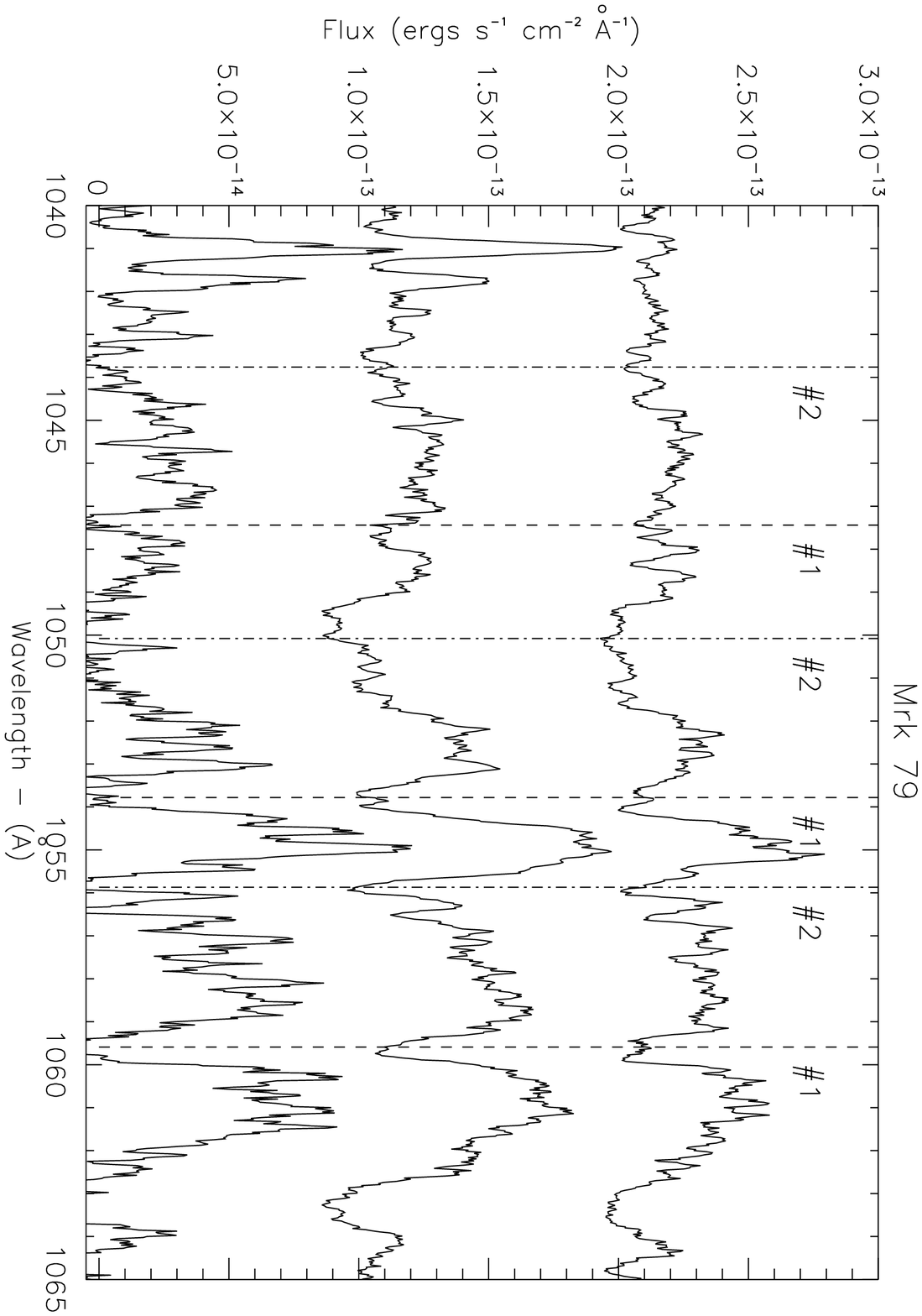]{The available spectra for Mrk 79 plotted in order of increasing time from
bottom to top. The upper spectra are offset by: 6.0$\times$10$^{-14}$ \& 1.2$\times$10$^{-13}$ ergs
s$^{-1}$ cm$^{-2}$ \AA$^{-1}$. Component 1 (as identified by Paper I)
is labeled by a dashed line and component 2 by a dotted and dashed line in each of the three lines 
(Ly$\beta$, O VIb and O VIr in order of increasing wavelength).}

\figcaption[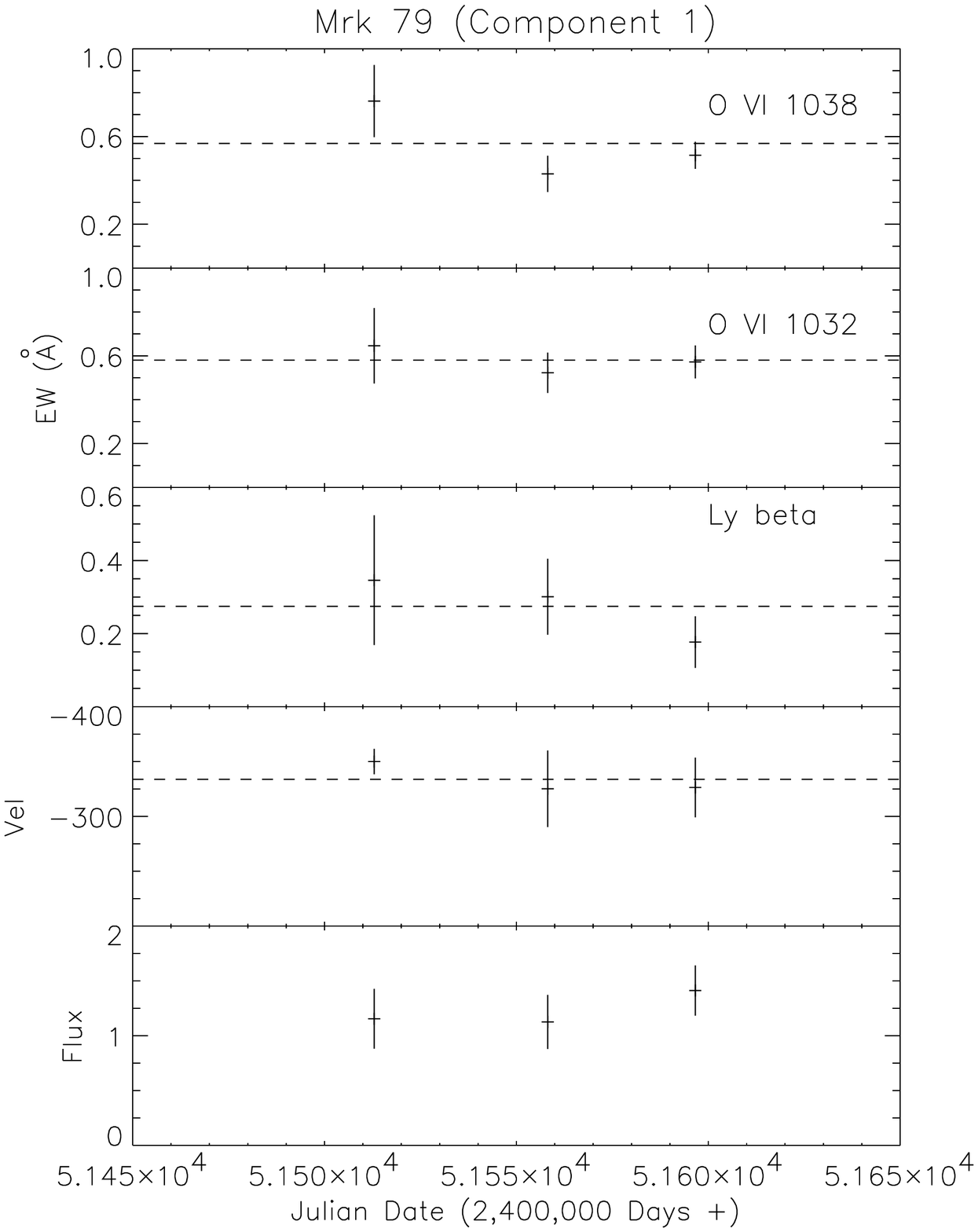]{This plot shows the relation for each of the three lines between EW, average 
velocity \&  flux over time. Top to bottom the lines are plotted in order of decreasing wavelength. The 
dashed line represents the average value line of the data, which is mostly to guide the eye. The continuum 
seems to increase slightly while Ly$\beta$ and O VI 1038 seem to vary.}

\figcaption[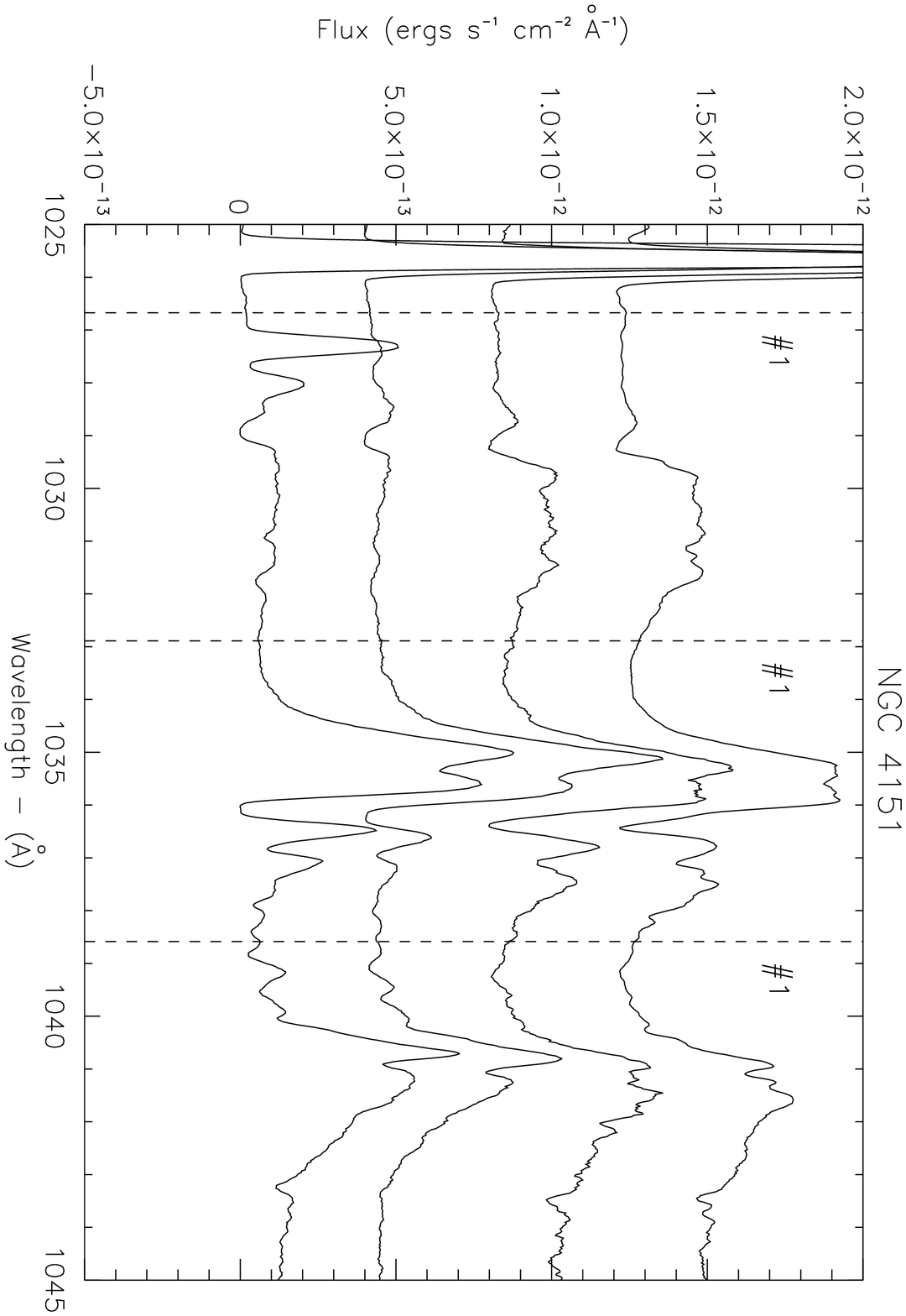]{Plotted in chronological order from bottom to top. The earlier spectra for NGC 4151 
show the flattening of the region between 1029 and 1032 \AA by an additional absorption component. 
The offsets for these data are: 4.0$\times$10$^{-13}$, 8.0$\times$10$^{-13}$ \& 1.2$\times$10$^{-12}$ ergs
s$^{-1}$ cm$^{-2}$ \AA$^{-1}$.}

\figcaption[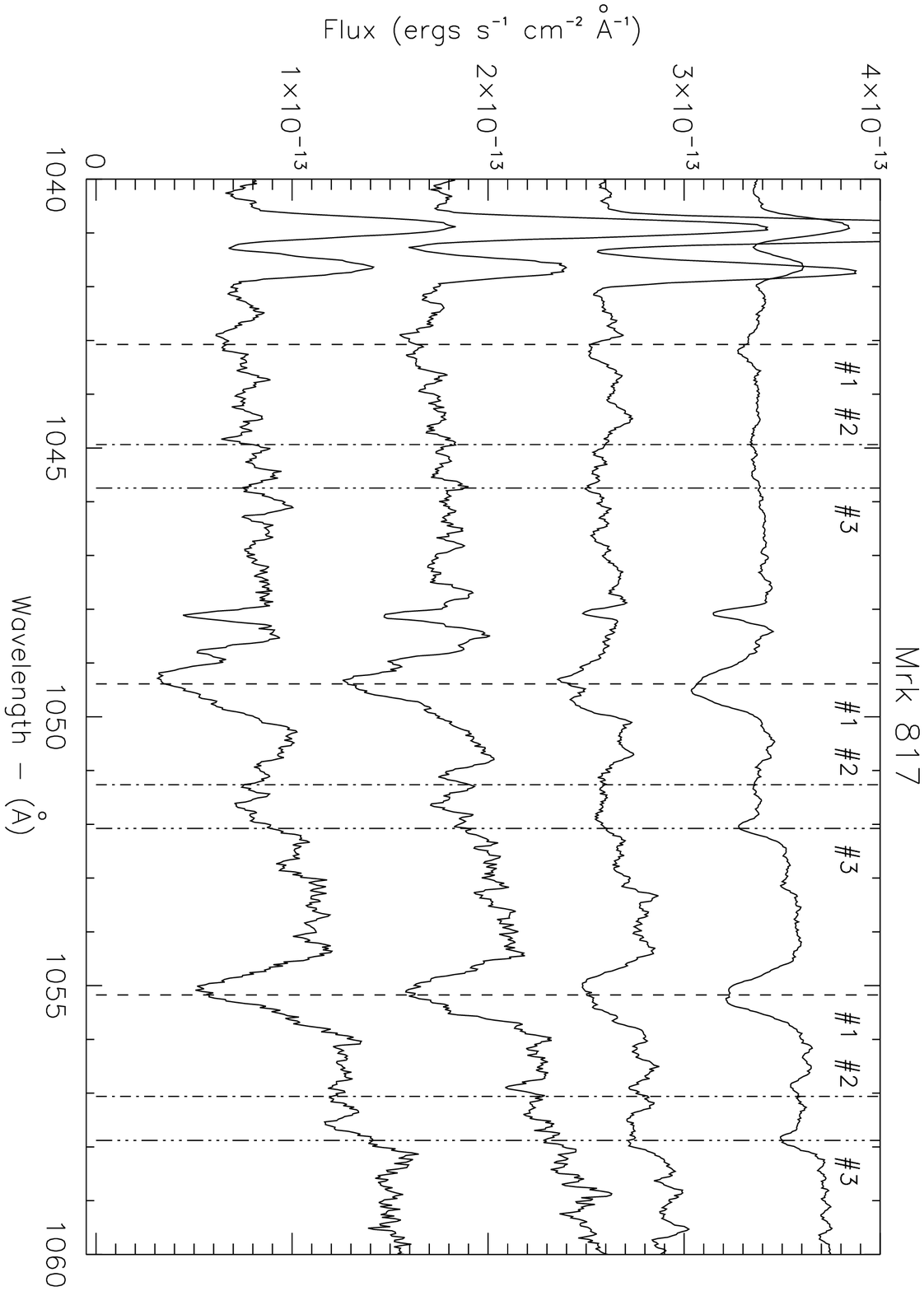]{These spectra show the changes seen in Mrk 817. We see variability in the O VIb 
line, and find absorption appearing between 1051 and 1052 \AA\ and 1056 and 1057 \AA (components 2 and 3). 
There is one point we were able to measure the continuum for, but the lines were lost in the S/N. We have
included the lightcurve point for completeness.}

\figcaption[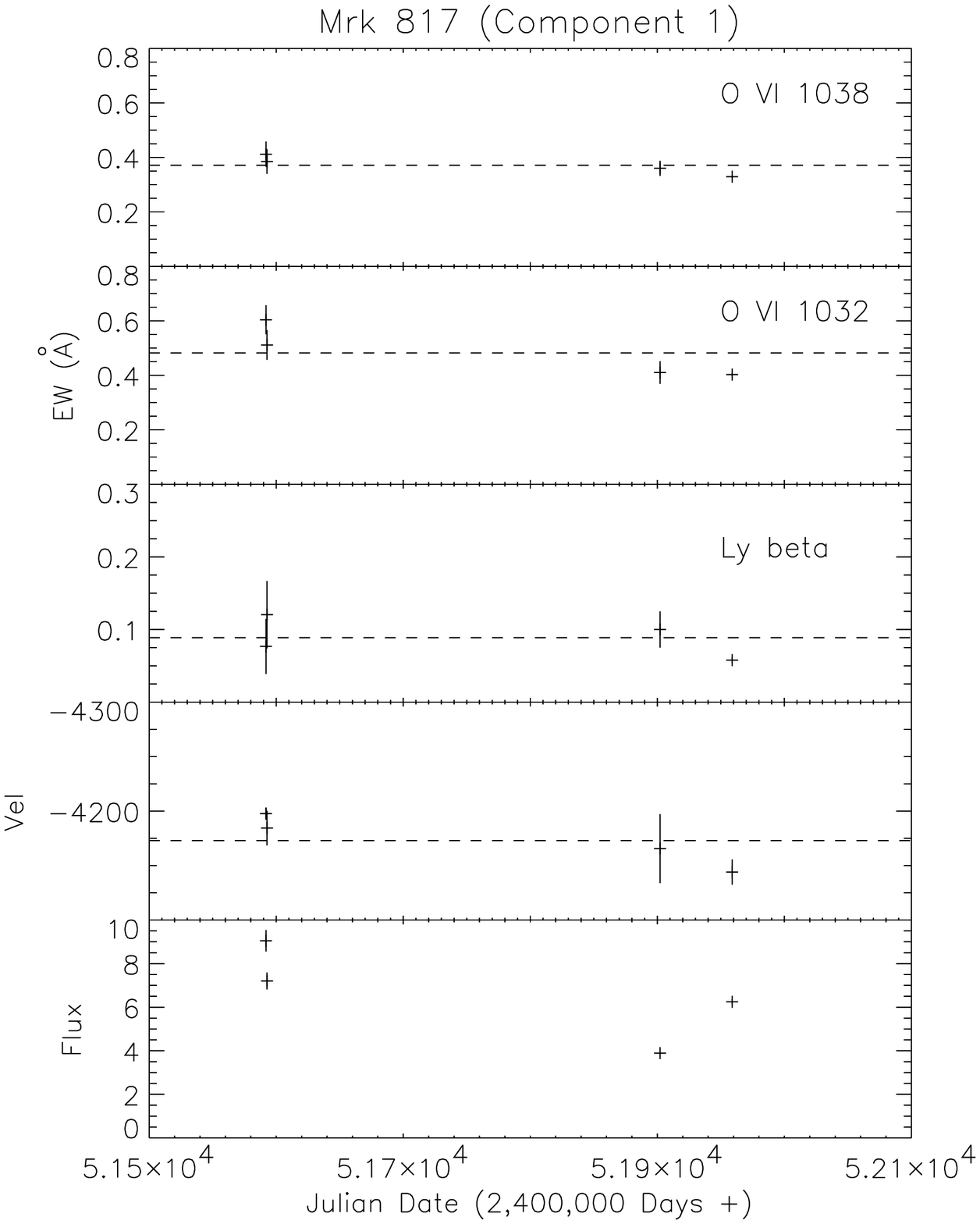]{Mrk 817 relationships of EW, velocity and flux for component 1over time. Both O VI 
lines show indications of decreasing EW with decreasing flux.}

\figcaption[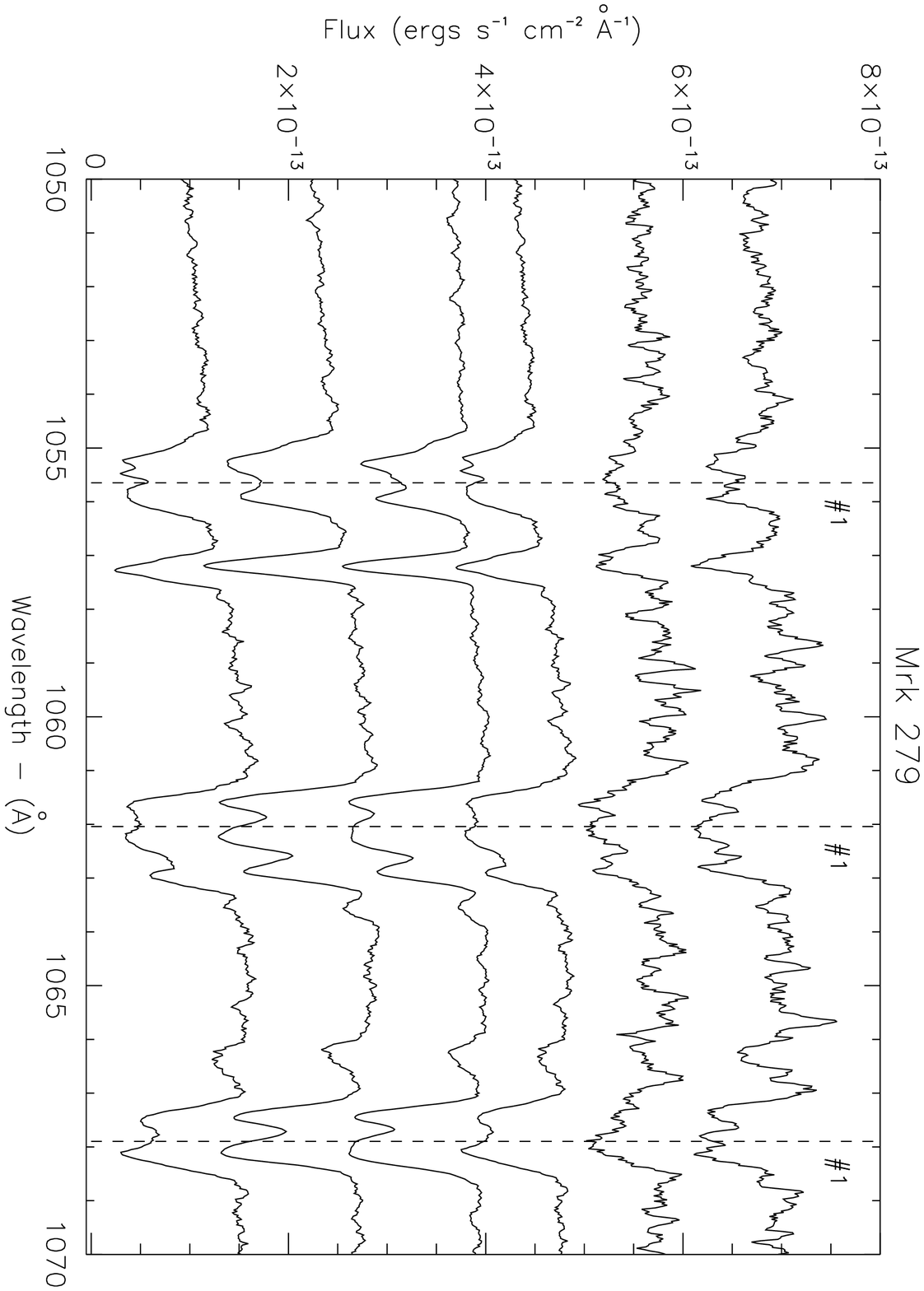]{We present the spectral time series for Mrk 279 with offsets in flux of: 1.0, 2.5, 
3.5, 5.0 \& 6.0 $\times$10$^{-13}$ ergs s$^{-1}$ cm$^{-2}$ \AA$^{-1}$. This shows the variability seen in 
the lines.}

\figcaption[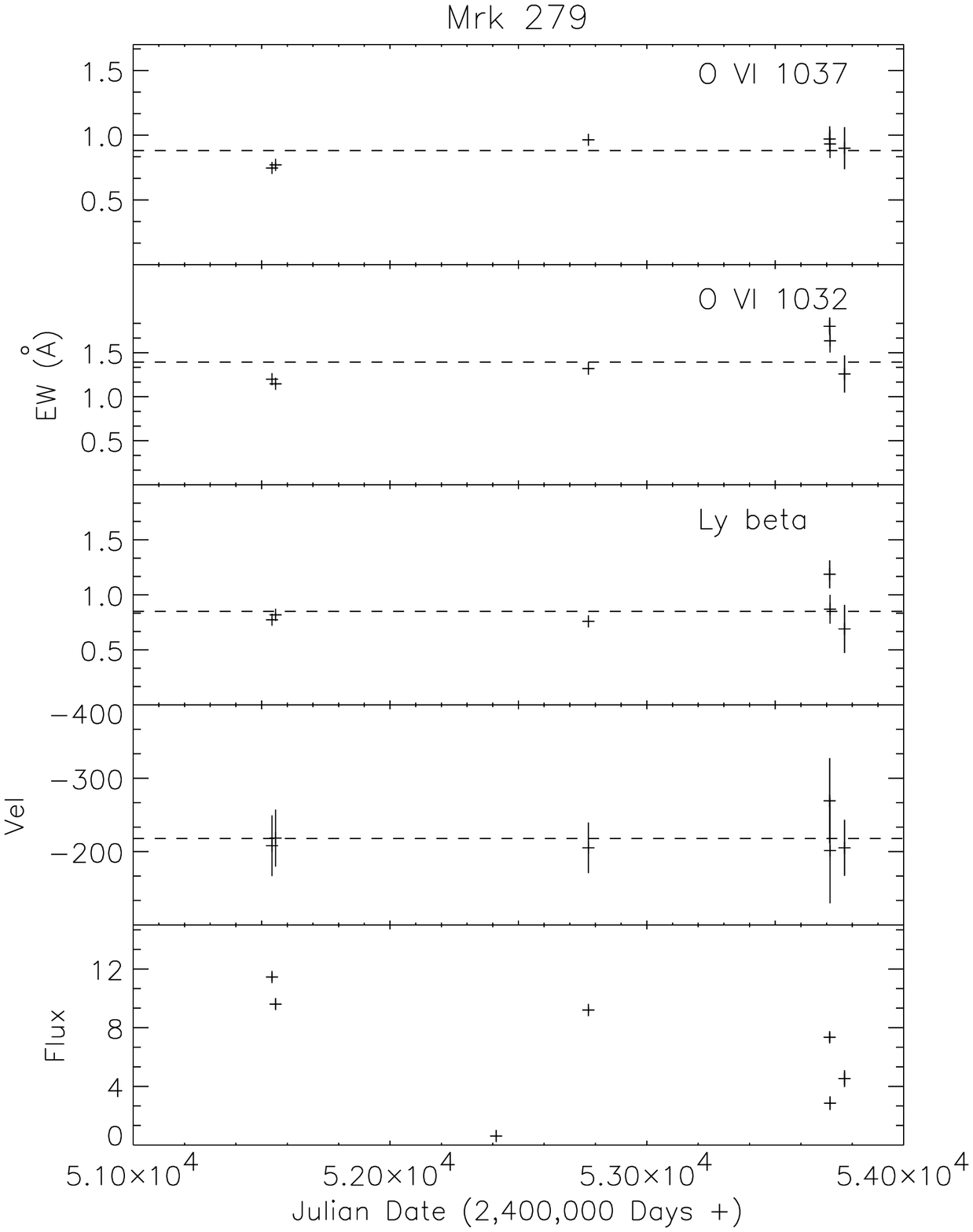]{Flux, EW and velocity against time for Mrk 279, similar to Figure 8. These quantities
exhibit definite 
variability with time, and appears to be slightly anticorrelated with the flux. The last three points
show signs of decrease in all three lines with the drop and rise of flux for those three observations,
which hints at a lag in response time.}

\figcaption[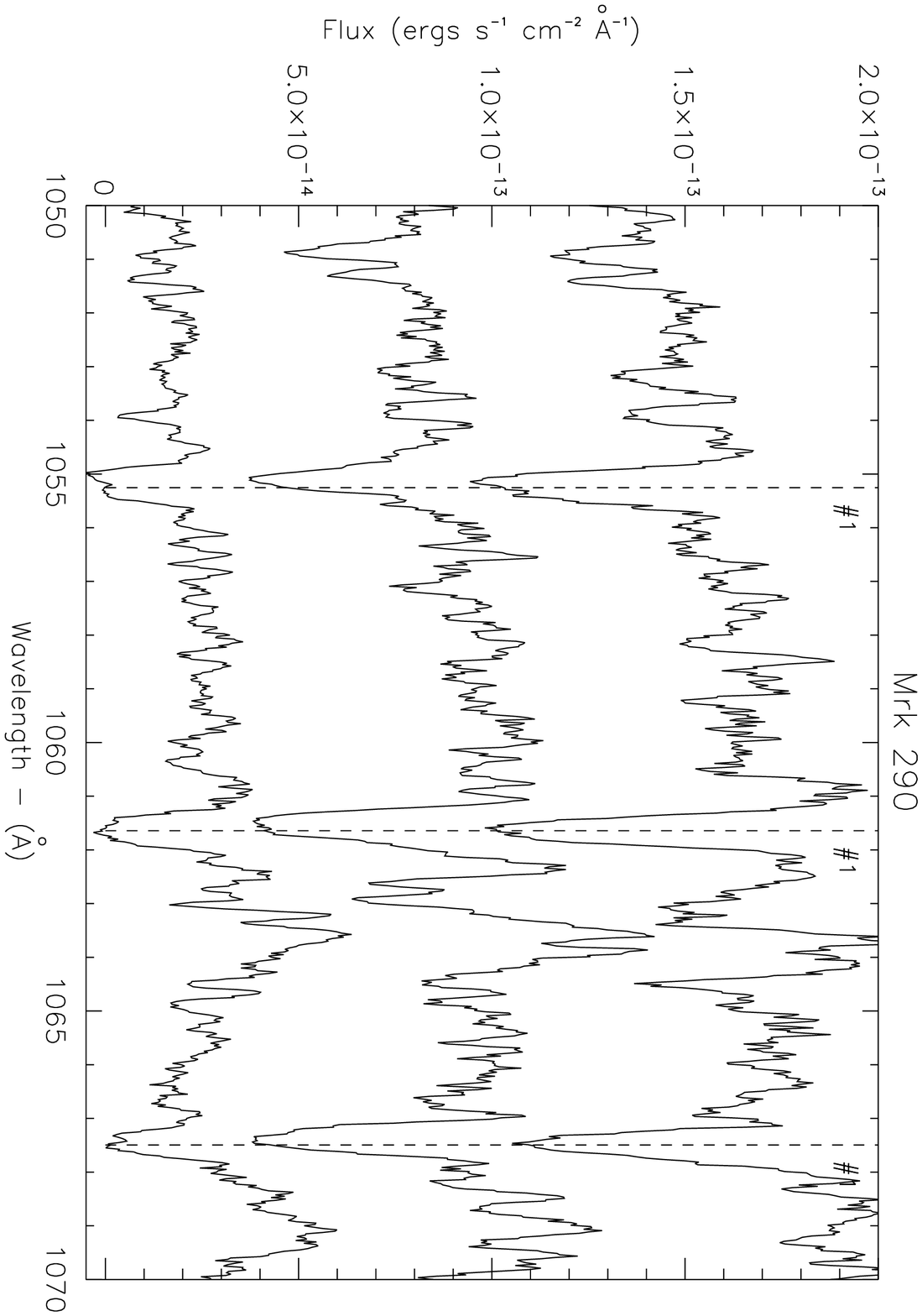]{These are spectra of Mrk 290, which show the line trends with time (bottom to top). 
The offsets for these spectra are: 3.0$\times$10$^{-14}$ \& 9.0$\times$10$^{-14}$ ergs
s$^{-1}$ cm$^{-2}$ \AA$^{-1}$.}

\figcaption[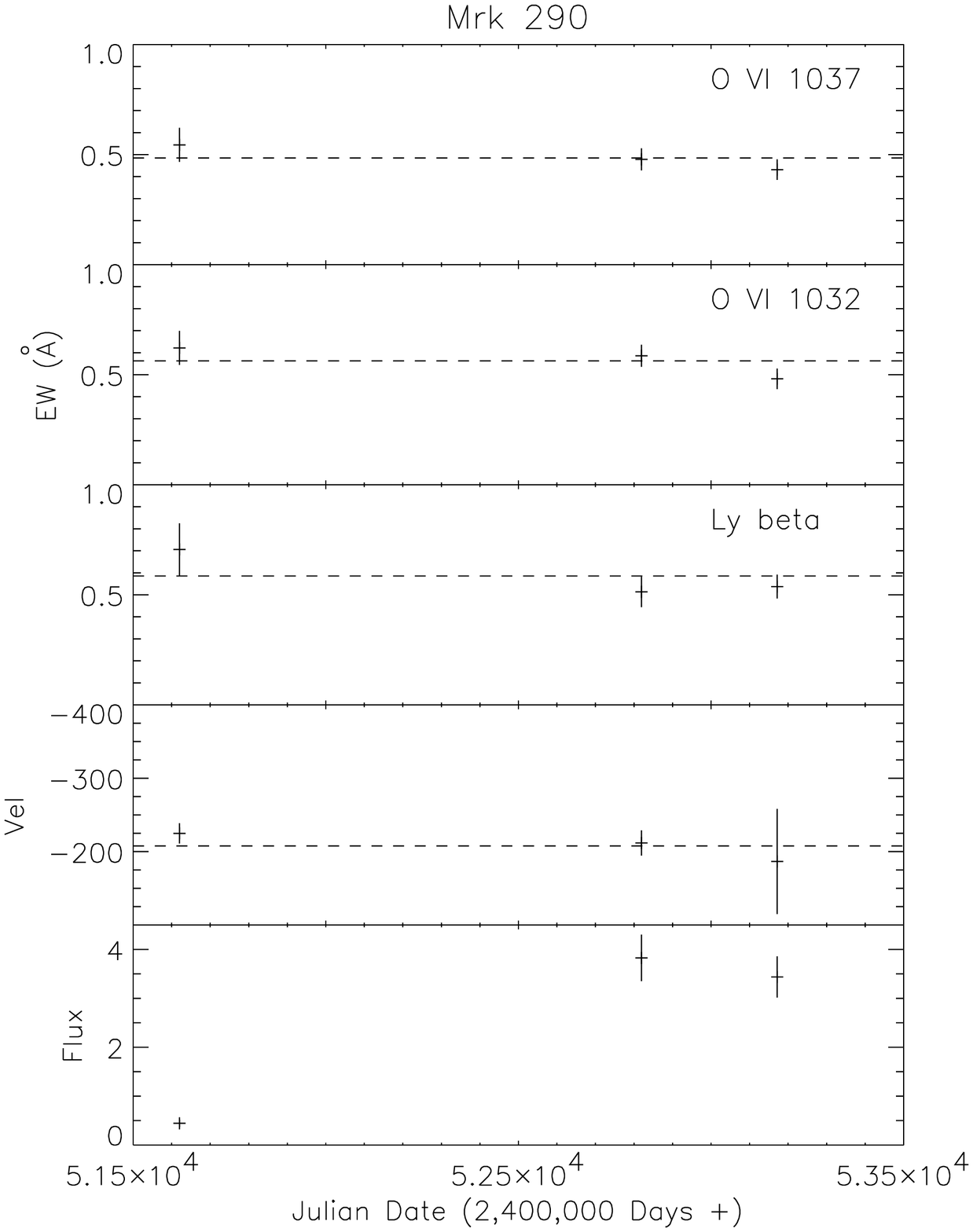]{We see in these plots (similar to Figure 8), that there is an decrease of EW with an 
increase in flux for all three lines. The trend for Lyman $\beta$ is less convincing, however, 
due to the larger error bars.}

\figcaption[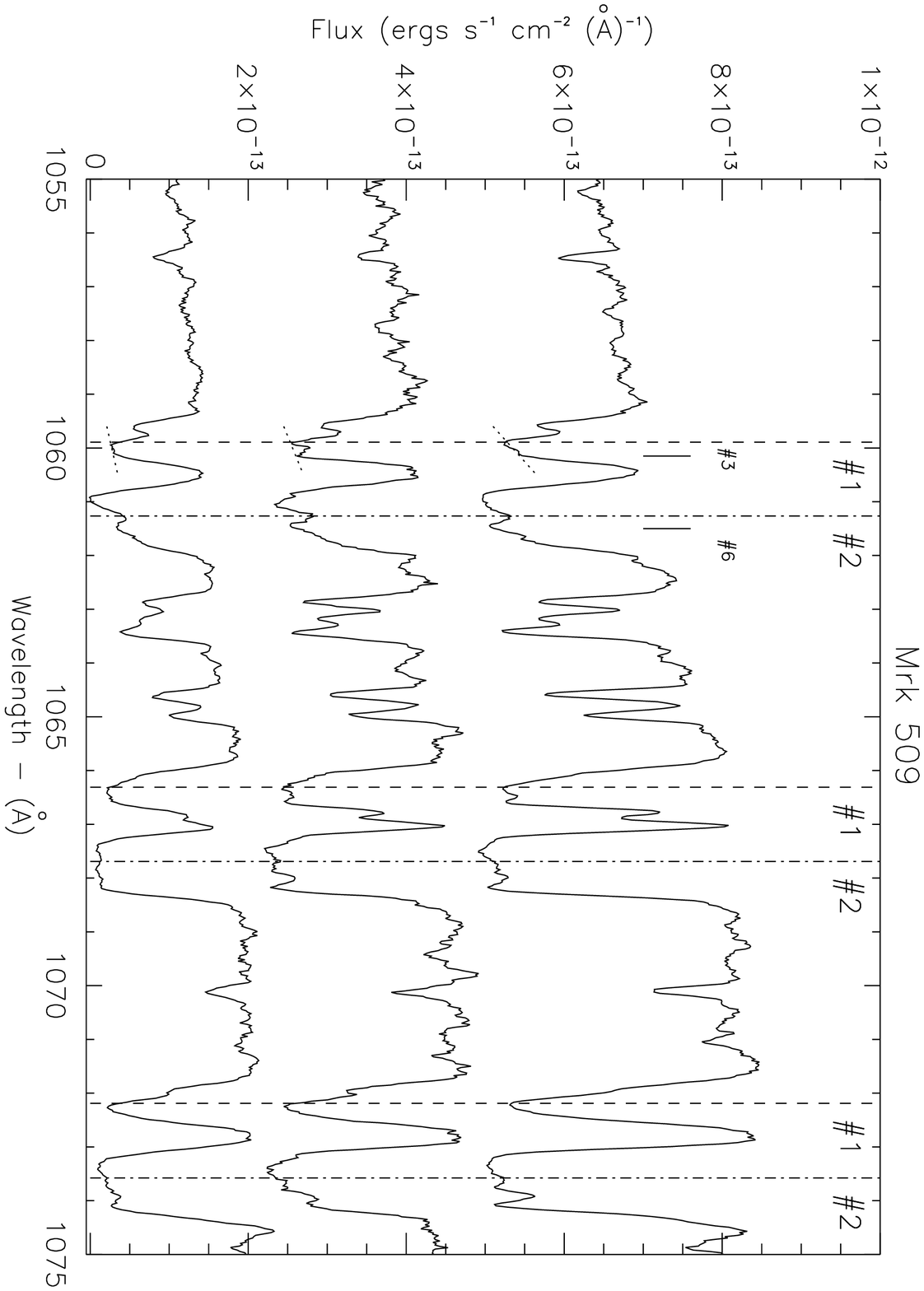]{We show the spectra for Mrk 509 and overplot the components 3 and 6 found by Kraemer 
et al. (2005), along with the dashed-dotted and dashed lines indicating our measured components 2 and 1 
respectively. We have placed dotted lines to show the change in the slope of component 3 with respect to 
component 2. The offsets for these are: 2.5 \& 5.0 $\times$10$^{-13}$ ergs
s$^{-1}$ cm$^{-2}$ \AA$^{-1}$.}

\figcaption[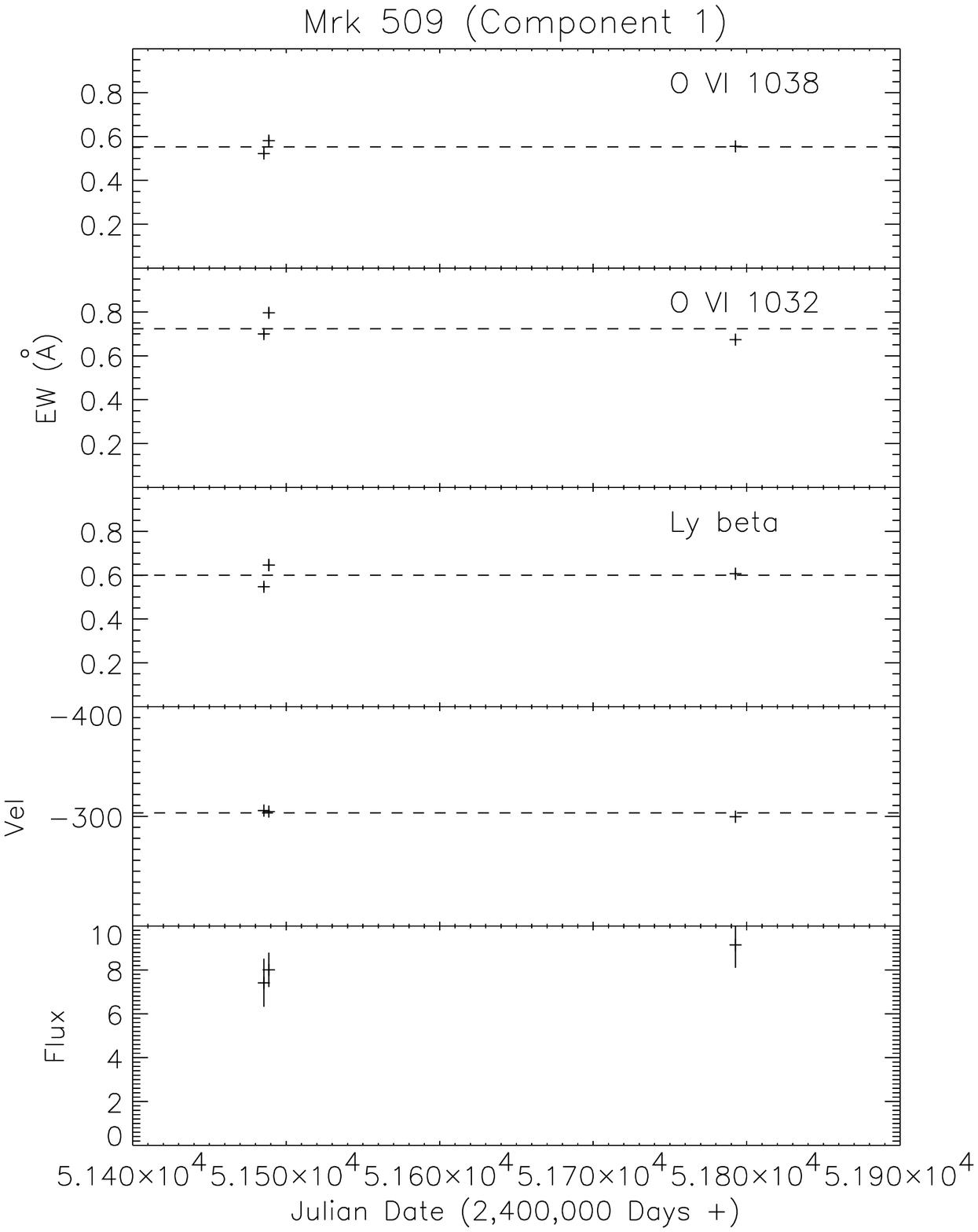]{Mrk 509 relations for component 1, plotted similar to Figure 8.}

\figcaption[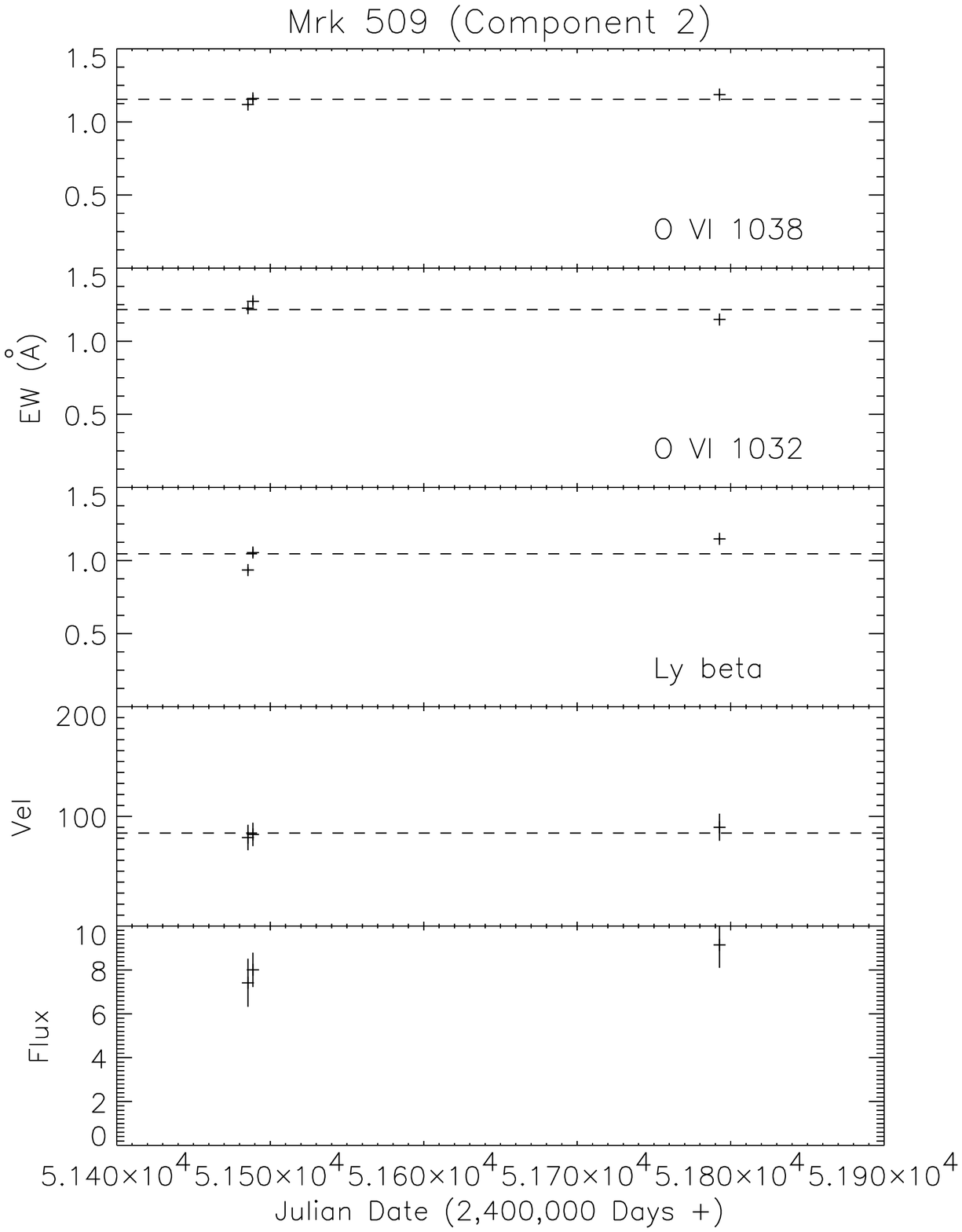]{Mrk 509 relations for component 2, plotted similar to Figure 8.}

\figcaption[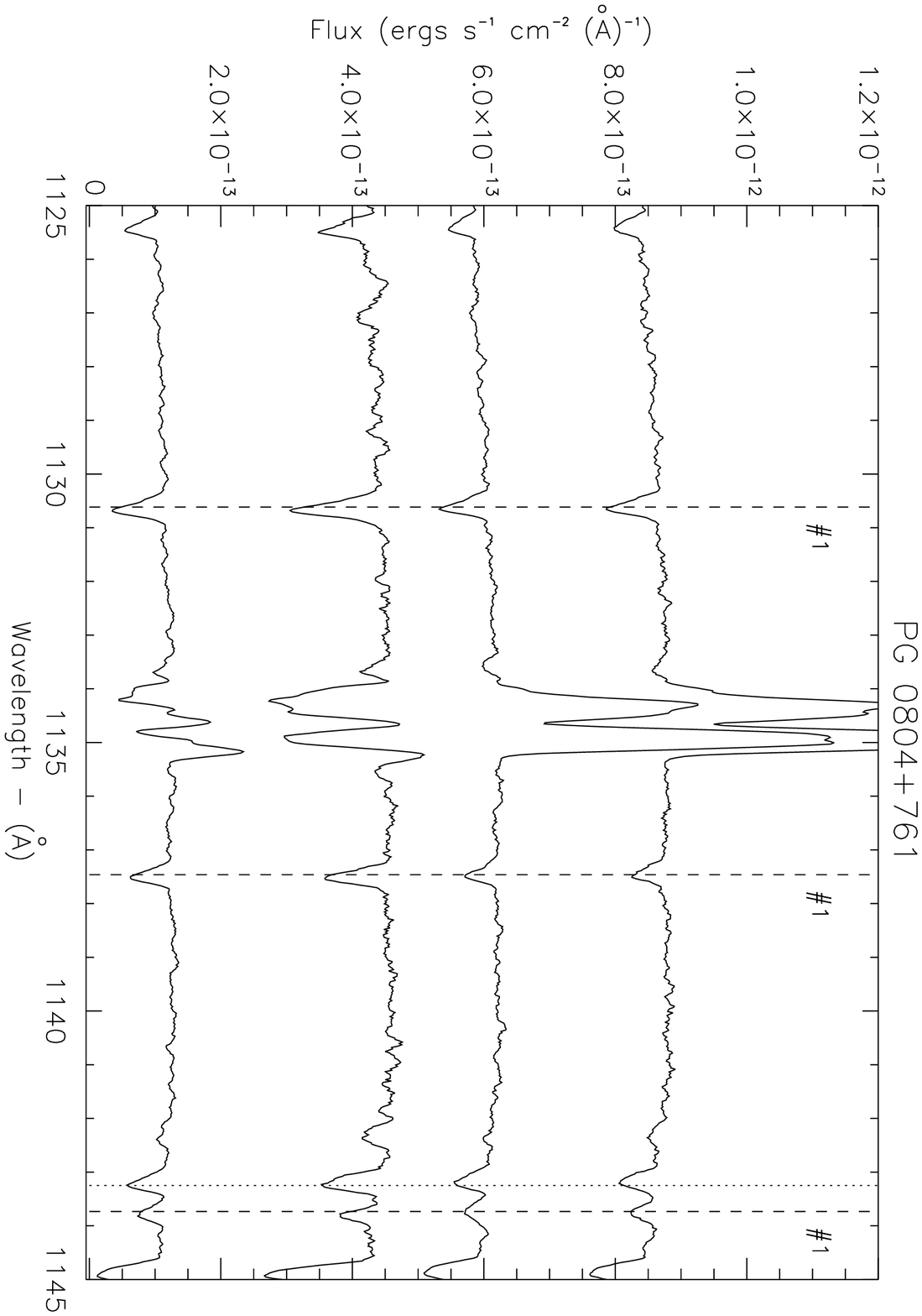]{Spectra for PG 0804+761. The lines are marked with the dashed average velocity line, 
and the lines clearly appear to move. The dotted line labels a galactic Fe II line $\lambda$1044.2 to 
show that the centroid for the ISM lines are not moving. Offsets for these are: 2.5, 5.0, 7.5 
$\times$10$^{-13}$ ergs s$^{-1}$ cm$^{-2}$ \AA$^{-1}$.}

\figcaption[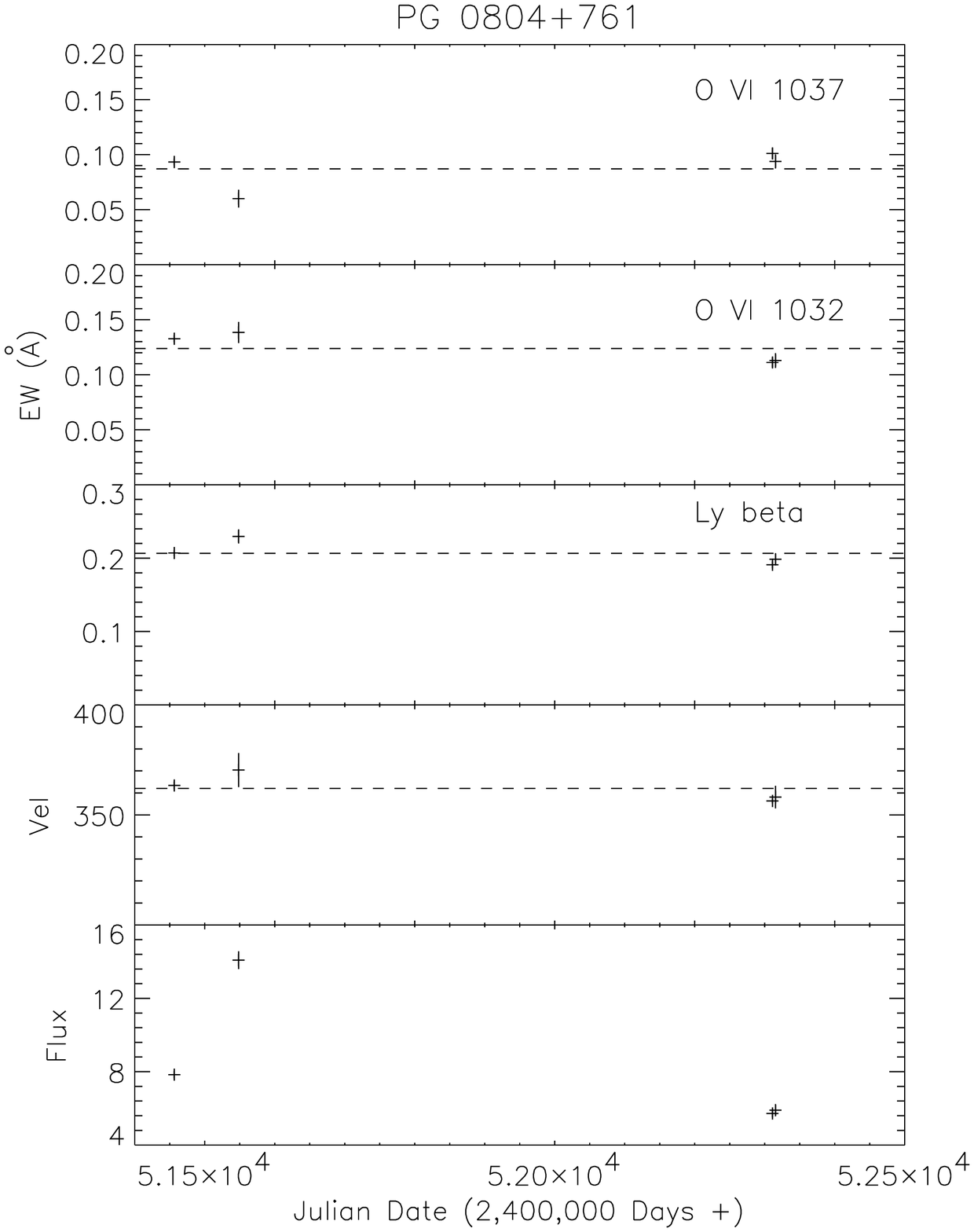]{PG 0804+761 relationships similar to Figure 8. We see definite correlations between 
velocity and EW with the flux levels with the exception of the Ly$\beta$ EW which seems to be low in the 
second observation.}

\figcaption[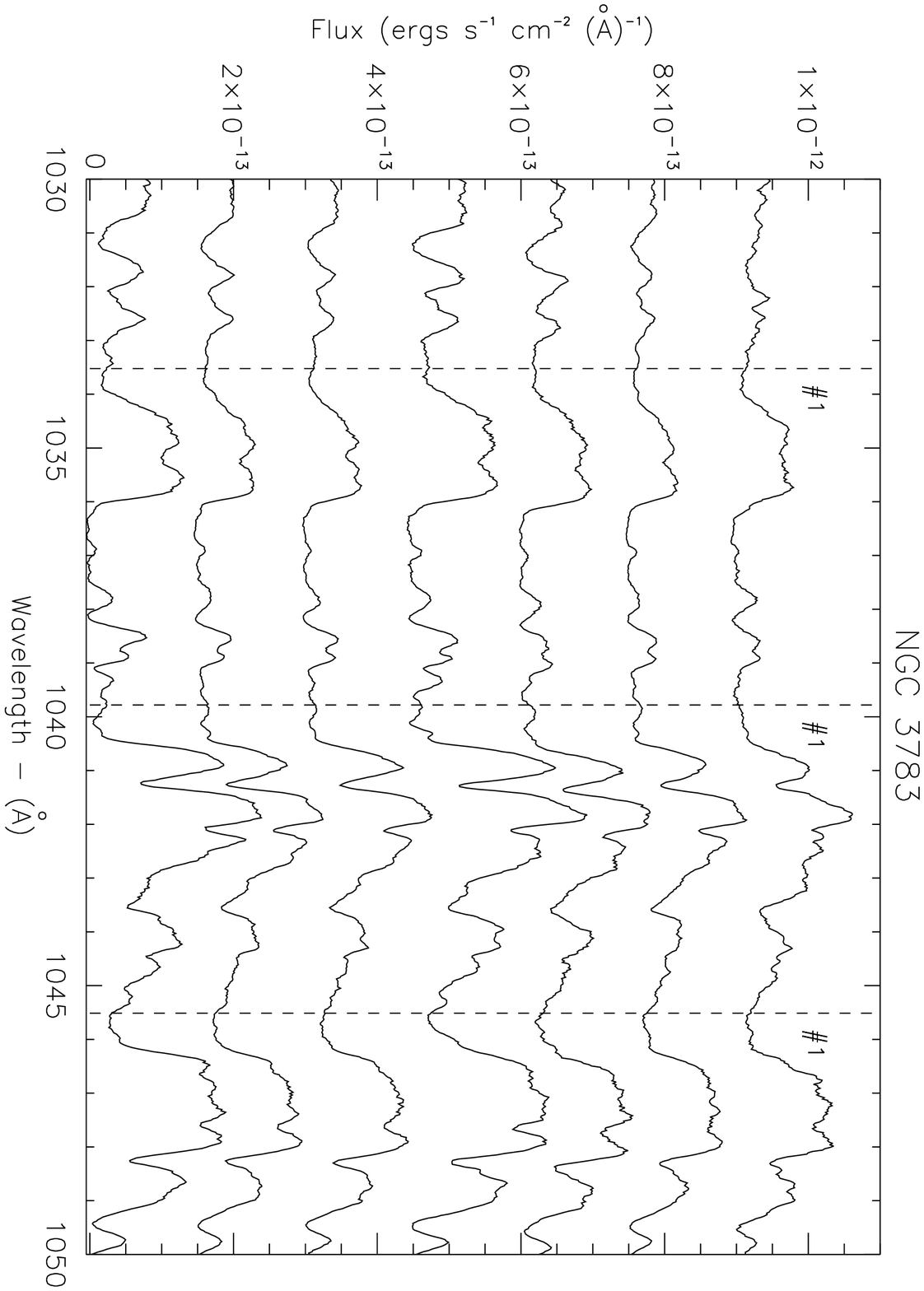]{Spectra showing the time evolution of NGC 3783. The measured component is labeled 
with the dashed line. Offsets for these are: 1.5, 3.0, 4.5, 6.0, 7.5 \& 9.0 $\times$10$^{-13}$ ergs
s$^{-1}$ cm$^{-2}$ \AA$^{-1}$.}

\figcaption[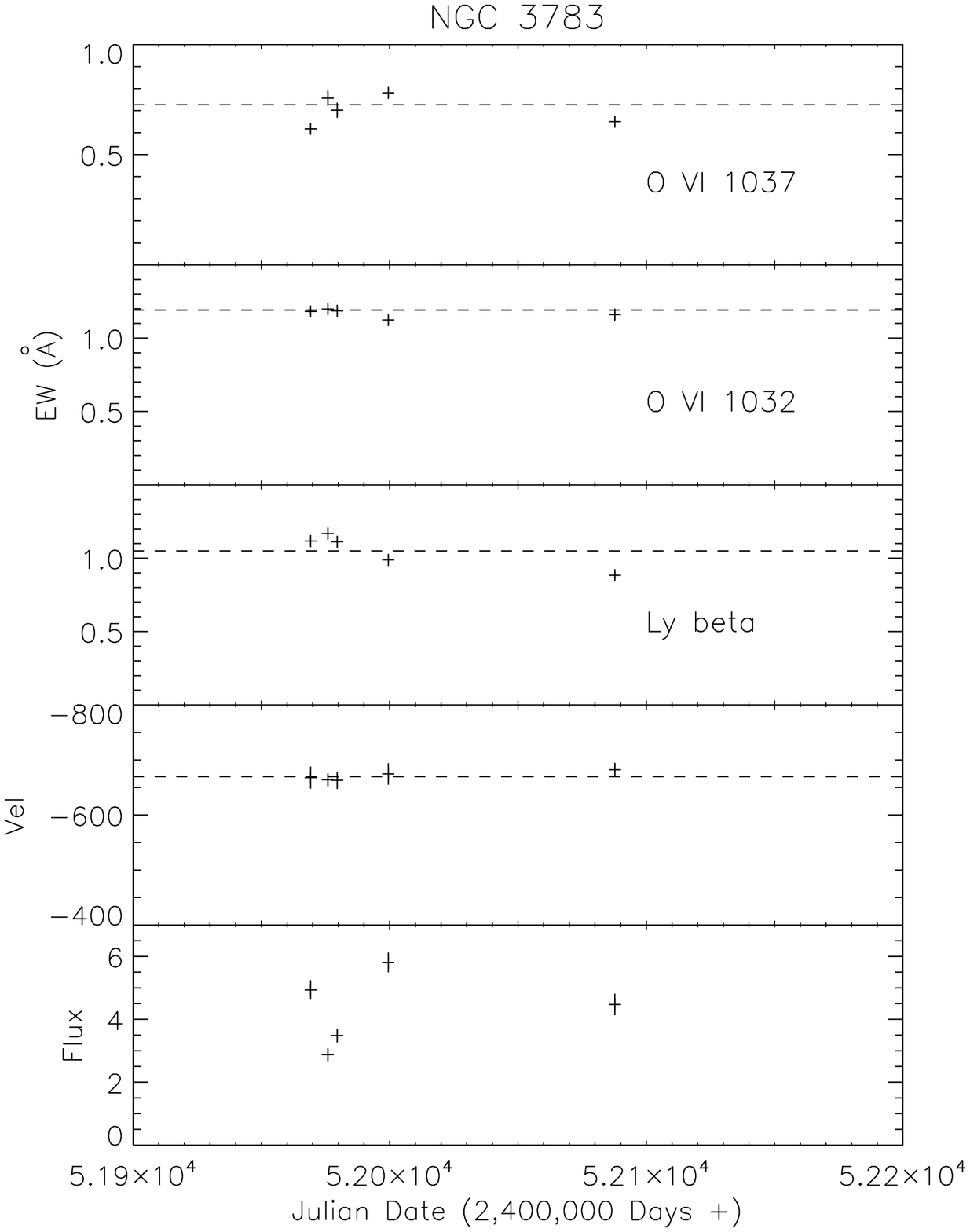]{This shows the relation for NGC 3783 between EW, velocity and flux over time. Similar
to Figure 8. We see definite variability in the EW over time for all three lines. There appears to be a
possible correlation between the change in Ly$\beta$ EW and continuum flux.}

\figcaption[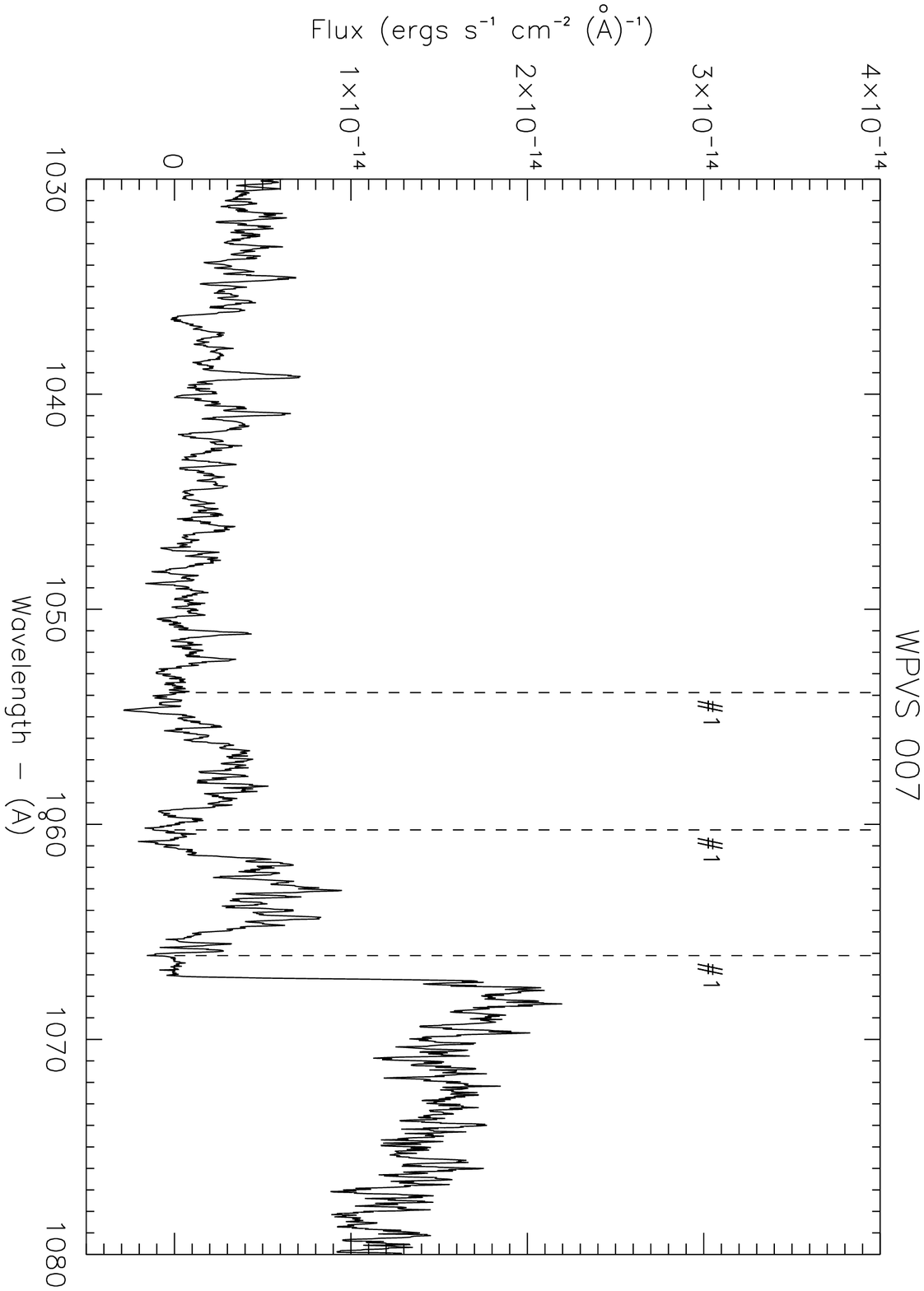]{Spectrum for WPVS 007. The three absorption features are marked with dashed lines.}

\clearpage
\begin{figure}
\plotone{Fig1.eps}
\\Fig.~1.
\end{figure}

\clearpage
\begin{figure}
\plotone{Fig2.eps}
\\Fig.~2.
\end{figure}

\clearpage
\begin{figure}
\plotone{Fig3.eps}
\\Fig.~3.
\end{figure}

\clearpage
\begin{figure}
\plotone{Fig4.eps}
\\Fig.~4.
\end{figure}

\clearpage
\begin{figure}
\plotone{Fig5.eps}
\\Fig.~5.
\end{figure}

\clearpage
\begin{figure}
\plotone{Fig6.eps}
\\Fig.~6a.
\end{figure}

\clearpage
\begin{figure}
\plotone{Fig7.eps}
\\Fig.~7.
\end{figure}

\clearpage
\begin{figure}
\plotone{Fig8.eps}
\\Fig.~8.
\end{figure}

\clearpage
\begin{figure}
\plotone{Fig9.eps}
\\Fig.~9.
\end{figure}

\clearpage
\begin{figure}
\plotone{Fig10.eps}
\\Fig.~10.
\end{figure}

\clearpage
\begin{figure}
\plotone{Fig11.eps}
\\Fig.~11.
\end{figure}

\clearpage
\begin{figure}
\plotone{Fig12.eps}
\\Fig.~12.
\end{figure}

\clearpage
\begin{figure}
\plotone{Fig13.eps}
\\Fig.~13.
\end{figure}

\clearpage
\begin{figure}
\plotone{Fig14.eps}
\\Fig.~14.
\end{figure}

\clearpage
\begin{figure}
\plotone{Fig15.eps}
\\Fig.~15.
\end{figure}

\clearpage
\begin{figure}
\plotone{Fig16.eps}
\\Fig.~16.
\end{figure}

\clearpage
\begin{figure}
\plotone{Fig17.eps}
\\Fig.~17.
\end{figure}

\clearpage
\begin{figure}
\plotone{Fig18.eps}
\\Fig.~18.
\end{figure}

\clearpage
\begin{figure}
\plotone{Fig19.eps}
\\Fig.~19.
\end{figure}

\clearpage
\begin{figure}
\plotone{Fig20.eps}
\\Fig.~20.
\end{figure}

\clearpage
\begin{figure}
\plotone{Fig21.eps}
\\Fig.~21.
\end{figure}

\clearpage
\begin{figure}
\plotone{Fig22.eps}
\\Fig.~22.
\end{figure}

\clearpage
\begin{figure}
\plotone{Fig23.eps}
\\Fig.~23.
\end{figure}

\newpage
\clearpage
\begin{deluxetable}{lcccc}
\tablecolumns{5}
\footnotesize
\tablecaption{Observation Log for CTIO Spectra}
\tablehead{
\colhead{Object} &
\colhead{Date} &
\colhead{Integration Time} &
\colhead{Resolution FWHM} &
\colhead{$\lambda$ Coverage} \\

\colhead{\footnotesize{ }} &
\colhead{\footnotesize{ }} &
\colhead{\footnotesize{(s)}} &
\colhead{\footnotesize{(\AA)}} &
\colhead{\footnotesize{(\AA)}} \\
}

\startdata
IRAS F22456-5125 & 2007 August 9 & 2400 & 4.3 & 3667-5409 \\
IRAS F22456-5125 & 2007 August 16 & 2400 & 3.1 & 5639-6947 \\
MR 2251-178 & 2007 August 15 & 2400 & 4.3 & 3667-5423 \\
MR 2251-178 & 2007 August 17 & 2400 & 3.1 & 5639-6947 \\
WPVS 007 & 2007 August 15 & 2400 & 4.3 & 3667-5423 \\
WPVS 007 & 2007 August 17 & 1800 & 3.1 & 5638-6947 \\

\enddata
\normalsize
\end{deluxetable}
\thispagestyle{empty}

\newpage
\begin{deluxetable}{lcccccc}
\tablecolumns{7}
\footnotesize
\tablecaption{AGN Continuum Levels}
\tablewidth{1.0\textwidth}
\tabletypesize{\scriptsize}
\tablehead{
\colhead{Object} &
\colhead{Observation ID} &
\colhead{Julian Date} &
\colhead{Flux} &
\colhead{Bin} &
\colhead{Log L$_{\lambda}$} &
\colhead{$\sigma$$_{Log L_{\lambda}}$}\\
\colhead{ } &
\colhead{ } &
\colhead{+2400000 days} &
\colhead{$\times$10$^{14}$} &
\colhead{(\AA)} &
\colhead{ergs s$^{-1}$ \AA $^{-1}$} &
\colhead{ergs s$^{-1}$ \AA $^{-1}$}\\
\colhead{ } &
\colhead{ } &
\colhead{ } &
\colhead{ergs s$^{-1}$ \AA $^{-1}$} &
\colhead{ } &
\colhead{ } &
\colhead{ }\\
}

\startdata
WPVS 007 & D8060201000 & 52950.360 & 0.48 & 1020 & 39.88 & 0.11 \\
QSO0045+3926  & Z0020401000  & 51874.176 & 1.01 & 1110 & 41.54 & 0.03 \\
 & D1310101000  & 52921.336 & 1.34 & 1110 & 41.66 & 0.05 \\
 & D1310102000  & 52922.707 & 1.40 & 1110 & 41.68 & 0.05 \\
 & D1310103000  & 52924.391 & 1.14 & 1110 & 41.60 & 0.04 \\
 & D1310104000  & 52983.242 & 0.88 & 1110 & 41.48 & 0.06 \\
 & D1310105000  & 53335.281 & 1.28 & 1110 & 41.65 & 0.07 \\
 & D1310106000  & 53336.105 & 1.24 & 1110 & 41.63 & 0.07 \\
 & D1310107000  & 53337.000 & 1.23 & 1110 & 41.63 & 0.07 \\
TONS180  & P1010502000  & 51524.742 & 6.16 & 1160 & 41.66 & 0.05 \\
 & D0280101000  & 53199.973 & 4.49 & 1160 & 41.52 & 0.05 \\
MRK1044  & D0410101000  & 53006.246 & 2.52 & 1020 & 40.12 & 0.09 \\
NGC985  & P1010903000  & 51536.551 & 2.45 & 1020 & 40.94 & 0.05 \\
EUVEJ0349-537  & E8970301000  & 53237.559 & 0.36 & 1110 & 41.06 & 0.04 \\
IRASF04250-5718  & D8080801000  & 52886.609 & 3.66 & 1110 & 41.89 & 0.07 \\
MRK79  & P1011701000  & 51512.949 & 1.15 & 1110 & 40.04 & 0.11 \\
 & P1011702000  & 51558.152 & 1.13 & 1110 & 40.03 & 0.10 \\
 & P1011703000  & 51596.613 & 1.41 & 1110 & 40.12 & 0.08 \\
MRK10  & Z9072801000  & 52673.195 & 1.07 & 1110 & 40.24 & 0.11 \\
IR07546+3928  & S6011801000  & 52317.062 & 0.11 & 1160 & 40.29 & 0.08 \\
PG0804+761  & P1011901000  & 51456.504 & 7.84 & 1160 & 42.18 & 0.01 \\
 & P1011903000  & 54548.312 & 14.08 & 1160 & 42.44 & 0.02 \\
 & S6011001000  & 52311.059 & 5.72 & 1160 & 42.05 & 0.02 \\
 & S6011002000  & 52315.406 & 5.90 & 1160 & 42.06 & 0.02 \\
Ton951  & P1012002000  & 53080.004 & 4.62 & 1160 & 41.56 & 0.06 \\
 & D0280304000  & 51595.316 & 3.42 & 1160 & 41.43 & 0.03 \\
 & D0280301000  & 53079.102 & 2.90 & 1160 & 41.36 & 0.07 \\
IRAS09149-62  & A0020503000  & 51581.129 & 1.42 & 1110 & 40.95 & 0.04 \\
 & S7011003000  & 53461.098 & 0.80 & 1110 & 40.70 & 0.09 \\
 & U1072202000  & 53929.031 & 0.74 & 1110 & 40.67 & 0.06 \\
MKN141  & D8061001000  & 53085.585 & 0.07 & 1110 & 39.09 & 0.38 \\
NGC3516  & P1110404000  & 51651.957 & 1.24 & 1020 & 38.97 & 0.18 \\
 & G9170101000  & 53776.217 & 0.69 & 1020 & 38.71 & 0.20 \\
 & G9170102000  & 54123.776 & 0.63 & 1020 & 38.67 & 0.23 \\
ESO265-G23 
 & A1210405000  & 51978.219 & 0.38 & 1110 & 40.36 & 0.13 \\
 & A1210407000  & 52341.223 & 0.39 & 1110 & 40.37 & 0.07 \\
 & A1210408000  & 52393.352 & 0.73 & 1110 & 40.64 & 0.03 \\
 & A1210409000  & 52394.416 & 1.03 & 1110 & 40.79 & 0.04 \\
NGC3783  & P1013301000  & 51576.621 & 4.90 & 1160 & 39.95 & 0.03 \\
 & B1070102000  & 51969.199 & 4.94 & 1160 & 39.95 & 0.03 \\
 & B1070103000  & 51979.535 & 3.49 & 1160 & 39.80 & 0.03 \\
 & B1070104000  & 51999.453 & 5.81 & 1160 & 40.02 & 0.03 \\
 & B1070105000  & 52087.727 & 4.47 & 1160 & 39.91 & 0.04 \\
 & B1070106000  & 51975.855 & 2.87 & 1160 & 39.72 & 0.03 \\
 & E0310101000  & 53130.965 & 4.11 & 1160 & 39.87 & 0.03 \\
NGC4051  & B0620201000  & 52362.582 & 1.29 & 1020 & 38.13 & 0.05 \\
 & C0190101000  & 52657.988 & 1.35 & 1020 & 38.15 & 0.07 \\
 & C0190102000  & 52717.824 & 1.51 & 1020 & 38.20 & 0.07 \\
NGC4151  & P1110505000  & 51609.410 & 2.50 & 1020 & 38.72 & 0.05 \\
 & P2110201000  & 52008.199 & 1.71 & 1020 & 38.56 & 0.08 \\
 & C0920101000  & 52423.469 & 21.52 & 1020 & 39.66 & 0.03 \\
 & P2110202000  & 52427.199 & 12.67 & 1020 & 39.43 & 0.05 \\
RXJ1230.8+0115  & P1019001000  & 51715.832 & 5.24 & 1110 & 42.14 & 0.06 \\
TOL1238-364  & D0100101000  & 51945.742 & 1.47 & 1110 & 39.53 & 0.06 \\
PG1351+640  & P1072501000  & 51562.117 & 1.19 & 1160 & 41.25 & 0.03 \\
 & S6010701000  & 52306.922 & 1.68 & 1160 & 41.40 & 0.02 \\
MRK279  & P1080303000  & 51540.527 & 11.45 & 1110 & 41.31 & 0.01 \\
 & P1080304000  & 51554.562 & 9.61 & 1110 & 41.23 & 0.02 \\
 & C0900201000  & 52413.258 & 0.61 & 1110 & 40.04 & 0.02 \\
 & D1540101000  & 52772.410 & 9.20 & 1110 & 41.21 & 0.01 \\
 & F3250103000  & 53711.961 & 7.35 & 1110 & 41.12 & 0.02 \\
 & F3250104000  & 53713.277 & 2.86 & 1110 & 40.70 & 0.08 \\
 & F3250106000  & 53769.598 & 4.53 & 1110 & 40.90 & 0.05 \\
RXJ135515+561244  & D8061601000  & 52712.387 & 0.39 & 1110 & 41.04 & 0.02 \\
PG1404+226  & P2100401000  & 52071.652 & 0.93 & 1110 & 41.23 & 0.10 \\
PG1411+442  & A0601010000  & 51675.637 & 2.01 & 1160 & 41.49 & 0.06 \\
NGC5548  & P1014601000  & 51703.055 & 1.55 & 1020 & 39.94 & 0.05 \\
MRK817  & P1080401000  & 51591.965 & 9.05 & 1110 & 41.23 & 0.03 \\
 & P1080402000  & 51592.766 & 7.20 & 1110 & 41.13 & 0.03 \\
 & P1080403000  & 51902.176 & 3.89 & 1110 & 40.87 & 0.003 \\
 & P1080404000  & 51958.891 & 6.25 & 1110 & 41.07 & 0.01 \\
MRK290  & P1072901000  & 51620.230 & 0.45 & 1110 & 39.87 & 0.08 \\
 & D0760101000  & 52819.457 & 3.44 & 1110 & 40.76 & 0.06 \\
 & E0840101000  & 53171.730 & 3.83 & 1110 & 40.81 & 0.06 \\
MRK876  & P1073101000  & 51467.766 & 0.24 & 1110 & 40.88 & 0.03 \\
 & D0280201000  & 52776.450 & 5.77 & 1110 & 42.26 & 0.08 \\
 & D0280203000  & 53049.797 & 5.71 & 1110 & 42.26 & 0.03 \\
MRK509  & X0170101000  & 51485.445 & 7.41 & 1020 & 41.23 & 0.05 \\
 & X0170102000  & 51488.703 & 8.00 & 1020 & 41.26 & 0.06 \\
 & P1080601000  & 51792.750 & 9.14 & 1020 & 41.32 & 0.05 \\
IIZW136  & P1018301000  & 51863.473 & 2.36 & 1160 & 41.25 & 0.04 \\
 & P1018302000  & 53152.684 & 3.72 & 1160 & 41.45 & 0.07 \\
 & P1018303000  & 53153.727 & 2.79 & 1160 & 41.33 & 0.07 \\
 & P1018304000  & 53310.586 & 3.71 & 1160 & 41.45 & 0.04 \\
AKN564  & B0620101000  & 52089.816 & 1.03 & 1110 & 40.08 & 0.02 \\
IRAS F22456-5125  & Z9073901000  & 52541.902 & 1.12 & 1110 & 41.33 & 0.10 \\
 & Z9073902000 & 52542.457 & 1.75 & 1110 & 41.52 & 0.04 \\
 & E8481401000  & 53194.629 & 6.08 & 1110 & 42.07 & 0.08 \\
MR 2251-178  & P1111010000  & 52081.012 & 0.99 & 1160 & 40.89 & 0.04 \\
NGC7469  & P1074101000  & 51724.101 & 1.30 & 1020 & 39.52 & 0.08 \\
 & C0900101000  & 52621.797 & 2.11 & 1020 & 39.73 & 0.14 \\
 & C0900102000  & 52621.767 & 2.17 & 1020 & 39.74 & 0.15 \\

\enddata
\normalsize
\end{deluxetable}
\thispagestyle{empty}

\newpage
\begin{deluxetable}{rcc|ccc|ccc|ccc|cc}
\tablecolumns{14}
\rotate
\tablewidth{1.2\textwidth}
\tabletypesize{\scriptsize}
\tablecaption{Measured Quantities}
\tablehead{
\multicolumn{1}{c}{\textbf{Object}} &
\multicolumn{1}{c}{\textbf{Obs ID}} &
\multicolumn{1}{c}{\textbf{Comp}} &
\multicolumn{3}{c}{\textbf{Equivalent Width}} &
\multicolumn{3}{c}{\textbf{$\sigma$$_{EW}$}} &
\multicolumn{3}{c}{\textbf{FWHM}} &
\multicolumn{1}{c}{\textbf{V$_r$}} &
\multicolumn{1}{c}{\textbf{$\sigma$$_{vel}$}}\\ 
\colhead{ } &
\colhead{ } &
\colhead{ } &
\colhead{Ly$\beta$} &
\colhead{O VIb} &
\colhead{O VIr} &
\colhead{Ly$\beta$} &
\colhead{O VIb} &
\colhead{O VIr} &
\colhead{Ly$\beta$} &
\colhead{O VIb} &
\colhead{O VIr} &
\colhead{ } &
\colhead{ }\\
\colhead{\footnotesize{ }} &
\colhead{\footnotesize{ }} &
\colhead{\footnotesize{ }} &
\colhead{\footnotesize{(\AA)}} &
\colhead{\footnotesize{(\AA)}} &
\colhead{\footnotesize{(\AA)}} &
\colhead{\footnotesize{(\AA)}} &
\colhead{\footnotesize{(\AA)}} &
\colhead{\footnotesize{(\AA)}} &
\multicolumn{3}{c}{(\footnotesize{km s$^{-1}$})} &
\colhead{\footnotesize{(km s$^{-1}$})} &
\colhead{\footnotesize{(km s$^{-1}$})}\\
}

\startdata
WPVS 007
 & D8060201000 & 1 & 1.91 & 2.06 & 1.76 & 0.14 & 0.13 & 0.14 & 714 & 639 & 615 & -397 & 26 \\
QSO0045+3926 
 & Z0020401000  & 1 & 0.29 & 0.61 & 0.29 & 0.04 & 0.06 & 0.05 & 100 & 172 & 69 & 361 & 13 \\
 & D1310101000  & 1 & 0.19 & 0.45 & 0.38 & 0.01 & 0.01 & 0.01 & 56 & 128 & 98 & 351 & 5 \\
 & D1310102000  & 1 & 0.19 & 0.38 & 0.29 & 0.02 & 0.02 & 0.02 & 69 & 101 & 107 & 345 & 12 \\
 & D1310103000  & 1 & 0.29 & 0.41 & 0.27 & 0.08 & 0.06 & 0.06 & 77 & 121 & 63 & 340 & 22 \\
 & D1310104000  & 1 & 0.17 & 0.36 & 0.31 & 0.01 & 0.02 & 0.02 & 82 & 107 & 74 & 346 & 16 \\
 & D1310105000  & 1 & 0.13 & 0.40 & 0.33 & 0.02 & 0.02 & 0.02 & 86 & 107 & 94 & 371 & 6 \\
 & D1310106000  & 1 & 0.16 & 0.42 & 0.28 & 0.02 & 0.02 & 0.02 & 69 & 91 & 85 & 375 & 7 \\
 & D1310107000  & 1 & 0.14 & 0.33 & 0.32 & 0.02 & 0.03 & 0.02 & 60 & 77 & 73 & 396 & 9 \\
TONS180 
 & P1010502000  & 1 &  &  & 0.13 &  &  & 0.03 &  &  & 65 & -1732 & \\
 & D0280101000  & 1 &  &  & 0.17 &  &  & 0.03 &  &  & 85 & -1792 & \\
MRK1044 
 & D0410101000  & 1 & 0.20 & 0.22 &  & 0.04 & 0.03 &  & 65 & 92 &  & -1110 & 2 \\
NGC985 
 & P1010903000  & 1 &  & 0.11 & 0.06 &  & 0.01 & 0.01 &  & 96 & 77 & -814 & 20 \\
 & P1010903000  & 2 & 0.12 & 0.26 & 0.20 & 0.02 & 0.01 & 0.01 & 34 & 72 & 55 & -678 & 6 \\
 & P1010903000  & 3 & 0.56 & 0.71 & 0.65 & 0.02 & 0.02 & 0.01 & 158 & 199 & 180 & -454 & 51 \\
 & P1010903000  & 4 & 0.15 & 0.16 & 0.23 & 0.02 & 0.01 & 0.01 & 41 & 44 & 63 & -268 & 17 \\
EUVEJ0349-537 
 & E8970301000  & 1 & 0.47 & 0.71 & 0.60 & 0.04 & 0.04 & 0.06 & 128 & 218 & 165 & 26 & 21 \\
IRASF04250-5718 
 & D8080801000  & 1 & 0.56 & 0.62 & 0.64 & 0.06 & 0.07 & 0.07 & 159 & 195 & 184 & -216 & 5 \\
 & D8080801000  & 2 & 0.15 & 0.25 & 0.20 & 0.06 & 0.03 & 0.03 & 41 & 66 & 51 & -64 & 2 \\
MRK79 
 & P1011701000  & 1 & 0.35 & 0.65 & 0.76 & 0.18 & 0.17 & 0.17 & 117 & 242 & 220 & -350 & 12 \\
 & P1011702000  & 1 & 0.30 & 0.52 & 0.43 & 0.10 & 0.09 & 0.08 & 175 & 240 & 160 & -325 & 35 \\
 & P1011703000  & 1 & 0.18 & 0.57 & 0.52 & 0.07 & 0.08 & 0.06 & 86 & 198 & 194 & -326 & 27 \\
 & P1011701000  & 2 &  &  & 0.39 &  &  & 0.13 &  &  & 255 & -1404 & \\
 & P1011702000  & 2 &  &  & 0.31 &  &  & 0.08 &  &  & 87 & -1387 & \\
 & P1011703000  & 2 &  &  & 0.28 &  &  & 0.07 &  &  & 81 & -1367 & \\
MRK10  & Z9072801000  & 1 & 1.03 &  & 0.92 & 0.09 &  & 0.13 & 344 &  & 227 & -126 & 18 \\
IR07546+3928 
 & S6011801000  & 1 & 1.07 & 1.49 & 1.00 & 0.05 & 0.03 & 0.02 & 280 & 436 & 351 & -1777 & 9 \\
 & S6011801000  & 2 & 0.98 & 1.48 & 1.37 & 0.03 & 0.03 & 0.03 & 263 & 395 & 361 & -1116 & 12 \\
PG0804+761 
 & P1011901000  & 1 & 0.21 & 0.13 & 0.09 & 0.004 & 0.003 & 0.003 & 69 & 67 & 64 & 363 & 7 \\
 & P1011903000  & 1 & 0.23 & 0.14 & 0.06 & 0.01 & 0.01 & 0.01 & 75 & 66 & 55 & 370 & 2 \\
 & S6011001000  & 1 & 0.19 & 0.11 & 0.10 & 0.01 & 0.004 & 0.01 & 63 & 64 & 81 & 356 & 1 \\
 & S6011002000  & 1 & 0.20 & 0.11 & 0.09 & 0.01 & 0.01 & 0.01 & 74 & 66 & 71 & 358 & 5 \\
TON951 
 & P1012002000  & 1 & 0.60 & 0.43 & 0.43 & 0.02 & 0.01 & 0.01 & 159 & 123 & 118 & 179 & 16 \\
 & D0280304000  & 1 & 0.69 & 0.42 & 0.46 & 0.05 & 0.03 & 0.02 & 190 & 150 & 128 & 173 & 25 \\
 & D0280301000  & 1 & 0.66 & 0.52 & 0.41 & 0.05 & 0.05 & 0.02 & 186 & 149 & 114 & 149 & 23 \\
IRAS09149-62 
 & A0020503000  & 1 &  & 1.72 & 1.26 &  & 0.22 & 0.12 &  & 504 & 389 & 18 & 18 \\
 & S7011003000  & 1 &  &  & 1.08 &  &  & 0.11 &  &  & 377 & 31 & \\
 & U1072202000  & 1 &  &  & 1.33 &  &  & 0.08 &  &  & 431 & 56 & \\
NGC3516 
 & P1110404000  & 1 & 0.27 & 1.76 & 0.06 & 0.07 & 0.09 & 0.03 & 168 & 510 & 52  & -1511 & 360 \\
 & G9170101000  & 1 & 0.33 & 1.40 & 0.14 & 0.13 & 0.17 & 0.20 & 177 & 468 & 133 & -1362 & 6.8 \\
 & G9170102000  & 1 & 0.23 & 0.53 & 0.15 & 0.07 & 0.07 & 0.04 & 133 & 202 & 63  & -1343 & 101 \\
 & P1110404000  & 2 &      & 0.54 & 0.08 &      & 0.06 & 0.05 &     & 250 & 162 & -866 & 125\\
 & G9170101000  & 2 & 0.38 & 0.57 & 0.64 & 0.16 & 0.12 & 0.20 & 212 & 239 & 331 & -888 & 38 \\
 & G9170102000  & 2 & 0.15 & 0.28 & 0.34 & 0.06 & 0.05 & 0.05 & 112 & 99  & 198 & -911 & 52 \\
 & P1110404000  & 3 &      & 0.28 & 0.35 &      & 0.04 & 0.03 &     & 123 & 46  & -410 & 15 \\
 & G9170101000  & 3 & 0.38 &      & 0.58 & 0.15 &      & 0.12 & 168 &     & 196 & -460 & 10 \\
 & G9170102000  & 3 & 0.34 & 0.32 & 0.38 & 0.06 & 0.06 & 0.05 & 113 & 98  & 100 & -456 & 2 \\
 & P1110404000  & 4 & 0.38 & 0.38 & 0.40 & 0.08 & 0.05 & 0.06 & 110 & 163 & 239 & -216 & 1  \\
 & G9170101000  & 4 & 0.58 &      & 0.46 & 0.17 &      & 0.23 & 226 &     & 174 & -236 & 30 \\
 & G9170102000  & 4 & 0.67 & 0.79 & 0.80 & 0.06 & 0.06 & 0.07 & 233 & 287 & 302 & -227 & 6  \\
ESO265-G23 
 & A1210405000  & 1 &  & 0.29 & 0.15 &  & 0.15 & 0.09 &  & 127 & 126 & -178 & 13 \\
 & A1210407000  & 1 &  & 0.34 & 0.32 &  & 0.05 & 0.04 &  & 141 & 73  & -157 & 5  \\
 & A1210408000  & 1 &  & 0.29 & 0.36 &  & 0.10 & 0.08 &  & 63  & 288 & -124 & 14 \\
 & A1210409000  & 1 &  & 0.34 & 0.30 &  & 0.13 & 0.10 &  & 92  & 81  & -157 & 26 \\
NGC3783 
 & P1013301000  & 1 & 1.03 & 1.19 & 0.78 & 0.02 & 0.02 & 0.03 & 403 & 463 & 404 & -660 & 10 \\
 & B1070102000  & 1 & 1.12 & 1.18 & 0.62 & 0.03 & 0.02 & 0.03 & 394 & 467 & 456 & -668 & 20 \\
 & B1070103000  & 1 & 1.11 & 1.19 & 0.70 & 0.04 & 0.03 & 0.03 & 400 & 452 & 449 & -663 & 16 \\
 & B1070104000  & 1 & 0.99 & 1.12 & 0.78 & 0.03 & 0.03 & 0.02 & 332 & 443 & 393 & -675 & 19 \\
 & B1070105000  & 1 & 0.88 & 1.16 & 0.65 & 0.03 & 0.03 & 0.03 & 377 & 459 & 404 & -682 & 13 \\
 & B1070106000  & 1 & 1.17 & 1.20 & 0.76 & 0.04 & 0.03 & 0.03 & 427 & 457 & 395 & -664 & 12 \\
 & E0310101000  & 1 & 1.05 & 1.30 & 0.80 & 0.02 & 0.03 & 0.02 & 380 & 467 & 371 & -657 & 30 \\
NGC4051 
 & B0620201000  & 1 &  & 2.28 &      &  & 0.00 &      &  & 672 &     & -374 & \\
 & C0190101000  & 1 &  & 2.13 &      &  & 0.04 &      &  & 696 &     & -348 & \\
 & C0190102000  & 1 &  & 2.14 & 1.88 &  & 0.05 & 0.04 &  & 670 & 627 & -359 & 12 \\
NGC4151 
 & P1110505000  & 1 &  & 0.91 &      &  & 0.04 &      &  & 473 &     & -706 & \\
 & P2110201000  & 1 &  & 1.48 & 1.21 &  & 0.07 & 0.04 &  & 514 & 522 & -574 & 222 \\
 & C0920101000  & 1 &  & 1.77 & 1.95 &  & 0.01 & 0.01 &  & 718 & 681 & -397 & 41 \\
 & P2110202000  & 1 &  & 1.66 & 1.99 &  & 0.08 & 0.05 &  & 662 & 716 & -413 & 56 \\
RXJ1230.8+0115 
 & P1019001000  & 1 &  & 1.90 & 1.57 &  & 0.09 & 0.08 &  & 542 & 476 & -2994 & 16 \\
 & P1019001000  & 2 &  & 0.27 & 0.26 &  & 0.06 & 0.05 &  & 71  & 69  & -2420 & 50 \\
 & P1019001000  & 3 &  & 1.19 & 1.07 &  & 0.08 & 0.06 &  & 311 & 278 & -2014 & 3 \\
 & P1019001000  & 4 &  & 1.17 & 1.08 &  & 0.06 & 0.08 &  & 304 & 280 &  120  & 2 \\
TOL1238-364 
 & D0100101000  & 1 &  & 1.10 & 1.51 &  & 0.14 & 0.15 &  & 339 & 454 & -252 & 75 \\
PG1351+640 
 & P1072501000  & 1 & 0.60 & 1.31 & 1.20 & 0.01 & 0.01 & 0.01 & 182 & 453 & 447 & -1705 & 73 \\
 & S6010701000  & 1 & 0.54 & 1.37 & 1.42 & 0.02 & 0.02 & 0.02 & 208 & 455 & 429 & -1715 & 60 \\
 & P1072501000  & 2 & 1.51 & 2.08 & 1.94 & 0.02 & 0.01 & 0.01 & 407 & 559 & 516 & -883 & 24 \\
 & S6010701000  & 2 & 1.23 & 2.16 & 2.12 & 0.02 & 0.02 & 0.02 & 333 & 579 & 565 & -899 & 11 \\
MRK279 
 & P1080303000  & 1 & 0.77 & 1.20 & 0.75 & 0.01 & 0.01 & 0.01 & 321 & 478 & 301 & -312 & 6  \\
 & P1080304000  & 1 & 0.82 & 1.15 & 0.77 & 0.02 & 0.02 & 0.02 & 328 & 482 & 307 & -328 & 4  \\
 & C0900201000  & 1 & 1.01 & 0.96 & 1.06 & 0.07 & 0.06 & 0.06 & 299 & 497 & 312 & -351 & 2  \\
 & D1540101000  & 1 & 0.76 & 1.32 & 0.97 & 0.01 & 0.01 & 0.01 & 290 & 474 & 319 & -308 & 8  \\
 & F3250103000  & 1 & 1.19 & 1.80 & 0.97 & 0.13 & 0.10 & 0.10 & 285 & 508 & 306 & -404 & 33 \\
 & F3250104000  & 1 & 0.87 & 1.64 & 0.93 & 0.13 & 0.13 & 0.11 & 392 & 518 & 366 & -302 & 12 \\
 & F3250106000  & 1 & 0.69 & 1.26 & 0.90 & 0.22 & 0.21 & 0.16 & 324 & 502 & 328 & -308 & 28 \\
RXJ135515+561244 
 & D8061601000  & 1 & 0.34 & 0.61 & 0.76 & 0.04 & 0.04 & 0.05 & 86 & 160 & 205 & -834 & 12 \\
 & D8061601000  & 2 &      & 0.69 & 0.48 &      & 0.05 & 0.03 &    & 180 & 186 & -163 & 28 \\
PG1404+226 
 & P2100401000  & 1 & 0.19 & 0.54 & 0.36 & 0.04 & 0.03 & 0.04 & 88  & 151 & 127 & -166 & 57 \\
 & P2100401000  & 2 &      & 1.47 & 1.11 &      & 0.06 & 0.07 &     & 330 & 284 &  134 & 17 \\
PG1411+442 
 & A0601010000  & 1 &  &  & 0.64 &  &  & 0.03 &  &  & 170 & 55 & \\
NGC5548 
 & P1014601000  & 1 &  & 0.35 & 0.25 &  & 0.02 & 0.02 &  & 161 & 195 & -441 & 4 \\
 & P1014601000  & 2 &  & 0.25 & 0.31 &  & 0.02 & 0.02 &  & 111 & 142 & -683 & 5 \\
MRK817 
 & P1080401000  & 1 & 0.08 & 0.60 & 0.41 & 0.05 & 0.04 & 0.05 & 319 & 207 & 143 & -4198 & 5 \\
 & P1080402000  & 1 & 0.12 & 0.51 & 0.39 & 0.05 & 0.05 & 0.06 & 236 & 237 & 146 & -4184 & 16 \\
 & P1080403000  & 1 & 0.10 & 0.41 & 0.36 & 0.01 & 0.02 & 0.02 & 134 & 273 & 265 & -4166 & 32 \\
 & P1080404000  & 1 & 0.06 & 0.40 & 0.33 & 0.01 & 0.00 & 0.01 & 210 & 208 & 154 & -4144 & 12 \\
 & P1080404000  & 2 &      & 0.12 & 0.05 &      & 0.01 & 0.01 &     & 243 & 142 & -3680 & 27 \\
 & P1080404000  & 3 &      & 0.12 & 0.09 &      & 0.01 & 0.01 &     & 118 & 139 & -2966 & 6  \\
MRK290 
 & P1072901000  & 1 & 0.71 & 0.62 & 0.54 & 0.12 & 0.10 & 0.08 & 185 & 190 & 142 & -225 & 14 \\
 & D0760101000  & 1 & 0.51 & 0.59 & 0.48 & 0.07 & 0.05 & 0.05 & 141 & 192 & 165 & -212 & 17 \\
 & E0840101000  & 1 & 0.54 & 0.48 & 0.43 & 0.05 & 0.05 & 0.05 & 180 & 157 & 158 & -187 & 72 \\
MRK876 
 & P1073101000  & 1 &  & 0.11 & 0.09 &  & 0.01 & 0.01 &  & 73  & 99 & -3735 & 3 \\
 & D0280203000  & 1 &  & 0.15 & 0.11 &  & 0.00 & 0.00 &  & 101 & 94 & -3724 & 0.8 \\
MRK509 
 & X0170101000  & 1 & 0.55 & 0.70 & 0.52 & 0.02 & 0.01 & 0.01 & 217 & 217 & 182 & -305 & 16 \\
 & X0170102000  & 1 & 0.65 & 0.80 & 0.58 & 0.03 & 0.02 & 0.02 & 218 & 204 & 190 & -304 & 19 \\
 & P1080601000  & 1 & 0.61 & 0.67 & 0.56 & 0.01 & 0.01 & 0.01 & 170 & 201 & 177 & -300 & 18 \\
 & X0170101000  & 2 & 0.95 & 1.23 & 1.12 & 0.02 & 0.01 & 0.01 & 303 & 329 & 346 & 103 & 12 \\
 & X0170102000  & 2 & 1.05 & 1.27 & 1.16 & 0.03 & 0.02 & 0.03 & 321 & 331 & 322 & 27 & 11 \\
 & P1080601000  & 2 & 1.15 & 1.15 & 1.19 & 0.01 & 0.01 & 0.01 & 330 & 333 & 341 & 90 & 12 \\
IIZW136 
 & P1018301000  & 1 &  & 0.42 &      &  & 0.02 &      &  & 146 &    & -1487 & \\
 & P1018302000  & 1 &  & 0.47 &      &  & 0.03 &      &  & 173 &    & -1512 & \\
 & P1018303000  & 1 &  & 0.49 &      &  & 0.04 &      &  & 96  &    & -1517 & \\
 & P1018304000  & 1 &  & 0.45 &      &  & 0.03 &      &  & 167 &    & -1512 & \\
 & P1018301000  & 2 &  &      & 0.28 &  &      & 0.02 &  &     & 87 & 9  & \\
 & P1018302000  & 2 &  &      & 0.30 &  &      & 0.04 &  &     & 96 & -2 & \\
 & P1018303000  & 2 &  &      & 0.18 &  &      & 0.03 &  &     & 61 & 9  & \\
 & P1018304000  & 2 &  &      & 0.31 &  &      & 0.02 &  &     & 99 & -7 & \\
AKN564 
 & B0620101000  & 1 & 1.64 & 1.28 & 1.31 & 0.04 & 0.02 & 0.03 & 528 & 304 & 280 & -65 & 10 \\
IRAS F22456-5125 
 & Z9073901000 & 1 & 0.06 & 0.32 & 0.28 & 0.04 & 0.06 & 0.06 & 42  & 110 & 105 & -787 & 11 \\
 & Z9073902000 & 1 & 0.07 & 0.19 & 0.17 & 0.01 & 0.01 & 0.01 & 81  & 64  & 69  & -803 & 22 \\
 & E8481401000 & 1 & 0.06 & 0.24 & 0.29 & 0.05 & 0.04 & 0.07 & 50  & 73  & 99  & -812 & 28 \\
 & Z9073901000 & 2 & 0.42 &      &      & 0.05 &      &      & 130 &     &     & -596 & \\
 & Z9073902000 & 2 & 0.35 &      &      & 0.01 &      &      & 97  &     &     & -612 & \\
 & E8481401000 & 2 & 0.44 &      &      & 0.06 &      &      & 110 &     &     & -617 & \\
 & Z9073901000 & 3 & 0.38 & 0.23 & 0.25 & 0.06 & 0.05 & 0.05 & 117 & 66  & 101 & -424 & 9 \\
 & Z9073902000 & 3 & 0.27 & 0.20 & 0.21 & 0.01 & 0.01 & 0.01 & 77  & 67  & 69  & -455 & 5 \\
 & E8481401000 & 3 & 0.33 & 0.38 & 0.26 & 0.05 & 0.06 & 0.06 & 96  & 151 & 94  & -447 & 13 \\
 & Z9073901000 & 4 & 0.09 & 0.17 & 0.10 & 0.05 & 0.05 & 0.04 & 56  & 66  & 42  & -311 & 3 \\
 & Z9073902000 & 4 & 0.08 & 0.21 & 0.14 & 0.01 & 0.01 & 0.01 & 40  & 63  & 66  & -323 & 5 \\
 & E8481401000 & 4 & 0.12 & 0.18 & 0.14 & 0.05 & 0.05 & 0.04 & 35  & 105 & 39  & -327 & 1 \\
 & Z9073901000 & 5 & 0.70 & 0.51 & 0.54 & 0.08 & 0.06 & 0.07 & 190 & 142 & 143 & -118 & 19 \\
 & Z9073902000 & 5 & 0.68 & 0.67 & 0.46 & 0.01 & 0.01 & 0.01 & 182 & 136 & 134 & -135 & 22 \\
 & E8481401000 & 5 & 0.72 & 0.54 & 0.57 & 0.07 & 0.06 & 0.07 & 178 & 147 & 137 & -136 & 21 \\
MR 2251-178 
 & P1111010000  & 1 & 1.62 & 0.70 & 0.40 & 0.03 & 0.01 & 0.01 & 418 & 234 & 155 & -300 & 23 \\
NGC7469 
 & C0900101000  & 1 &  & 0.40 & 0.30 &  & 0.002 & 0.01 &  & 151 & 137 & -1847 & 17 \\
 & C0900102000  & 1 &  & 0.48 & 0.40 &  & 0.01 & 0.01 &  & 209 & 190 & -1878 & 3 \\
\enddata

\end{deluxetable}

\newpage
\begin{deluxetable}{lcccccc}
\tablecolumns{7}
\footnotesize
\tablecaption{Central Black Hole Masses}
\tablehead{
\colhead{Object} &
\colhead{FWHM H$\beta$} &
\colhead{Mass} &
\colhead{log $\lambda$L$_{\lambda}$} &
\colhead{log L$_{Bol}$} &
\colhead{L/L$_{Edd}$} &
\colhead{reference$^b$}\\

\colhead{\footnotesize{ }} &
\colhead{\footnotesize{km s$^{-1}$}} &
\colhead{\footnotesize{$\times$10$^{6}$}} &
\colhead{\footnotesize{ergs s$^{-1}$}} &
\colhead{\footnotesize{ergs s$^{-1}$}} &
\colhead{\footnotesize{ }} &
\colhead{\footnotesize{ }}\\
}

\startdata
Mrk 79  &  & 52.4 & 43.72 & 44.71 & 0.08 & 1 \\
PG 0804+761 &  & 693 & 44.94 & 45.93 & 0.10 & 1 \\
NGC 3516 &  & 42.7 & 42.88 & 43.87 & 0.01 & 1 \\
NGC 3783 &  & 29.8 & 43.26 & 44.25 & 0.05 & 1 \\
NGC 4051 &  & 1.91 & 41.93 & 42.92 & 0.03 & 1 \\
NGC 4151 &  & 13.3 & 42.88 & 43.87 & 0.04 & 1 \\
Mrk 279 &  & 34.9 & 43.88 & 44.87 & 0.17 & 1 \\
PG 1411+422 &  & 443 & 44.63 & 45.62 & 0.07 & 1 \\
NGC5548 &  & 67.1 & 43.51 & 44.50 & 0.04 & 1 \\
Mrk 817 &  & 49.4 & 43.82 & 44.81 & 0.10 & 1 \\
Mrk 509 &  & 143 & 44.28 & 45.27 & 0.10 & 1 \\
NGC 7469 &  & 12.2 & 43.72 & 44.71 & 0.33 & 1 \\
Mrk 876 &  & 279 & 44.98 & 45.97 & 0.27 & 1 \\
Ton 951 &  & 92.4 & 44.35 & 45.34 & 0.19 & 1 \\
II Zw 136 &  & 457 & 44.46 & 45.45 & 0.05 & 1 \\
PG 1351+640 & 1170 & 46 & 44.64 & 45.63 & 0.74 & 2 \\
QSO 0045+3926 & 5970 & 564.9 & 44.60 & 45.59 & 0.05 & 7 \\
Ton S180 & 970 & 10.2 & 44.27 & 45.26 & 1.42 & 3 \\
NGC 985 & 5670 & 202.8 & 43.80 & 44.79 & 0.02 & 3 \\
IRAS F04250-5718 & 2580 & 144.0 & 44.87 & 45.86 & 0.40 & 3 \\
Mrk 10 & 3050 & 25.6 & 43.08 & 44.07 & 0.04 & 4 \\
IR 07546+3928 & 2120 & 176.9 & 45.39$^a$ & 46.39 & 1.10 & 8 \\
Mrk 141 & 3600 & 65.0 & 43.60 & 44.59 & 0.05 & 3 \\
RXJ 135515+561244 & 1100 & 9.4 & 43.98 & 44.97 & 0.79 & 3 \\
PG 1404+226 & 790 & 7.7 & 44.38 & 44.37 & 2.43 & 5 \\
Mrk 290 & 5320 & 157.3 & 43.69 & 44.68 & 0.02 & 5 \\
Akn 564 & 970 & 4.8 & 43.62 & 44.61 & 0.67 & 4 \\
MR 2251-5125 & 3768 & 240.0 & 44.64 & 45.63 & 0.14 & 6 \\
IRAS F22456-5125 & 3297 & 154.9 & 42.76$^a$ & 45.53 & 0.16 & 6 \\
WPVS 007 & 1502 & 11.5 & 43.59 & 44.58 & 0.27 & 6 \\
RXJ 1230.8+0115 &  &  &  & 45.0 &  0.13 & 9 \\

\enddata
\tablenotetext{a}{Mass calculated from H$\beta$ luminosity}
\tablenotetext{b}{1-Peterson et al. 2004, 2-Kaspi et al. 2000, 3-Grupe et al. 2004, 4-Botte et al. 2004, 
5-Vestergaard et al. 2006 6-This Paper, 7-Xu et al. 2003, 8-Marziani et al. 2003, 9-Ganguly et al. 2003}
\tablenotetext{c}{L$_{Bol}$ = 9.8 $\lambda$L$_{5100}$}
\normalsize
\end{deluxetable}
\thispagestyle{empty}

\end{document}